\documentclass[fleqn,usenatbib]{mnras}
\usepackage{graphicx}
\usepackage{natbib}
\usepackage{epstopdf}
\usepackage{amssymb}
\usepackage{amsmath}
\usepackage{mathptmx}
\usepackage{multirow}
\usepackage{array,booktabs}
\usepackage{hyperref}
\usepackage{subfig}
\usepackage[normalem]{ulem}
\usepackage{color,soul}
\usepackage[export]{adjustbox}
\usepackage{svg}
\usepackage[utf8]{inputenc}
\usepackage{appendix}
\usepackage{float}
\usepackage{caption}
\usepackage{comment}

\title[Fission and fusion in BNS mergers by RMS]{Fission and fusion of heavy nuclei induced by the passage of a radiation-mediated shock in BNS mergers}
\author[Granot et al.]{Alon Granot\thanks{E-mail: alongranot1@mail.tau.ac.il}, Amir Levinson, Ehud Nakar
\\
 School of Physics and Astronomy, Tel Aviv University, Tel Aviv 69978, Israel\\
 }
\date{May 2024}

\graphicspath{{figures/}}

\begin{document}
\label{firstpage}
\pagerange{\pageref{firstpage}--\pageref{lastpage}}
\maketitle

\begin{abstract}

We compute the structure of a Newtonian, multi-ion radiation-mediated shock (RMS) for different compositions anticipated in various stellar explosions. We use a multifluid  RMS model that incorporates electrostatic coupling between the different plasma constituents as well as Coulomb friction in a self-consistent manner, and approximates the effect of pair creation and the presence of free neutrons in the shock upstream on
the shock structure.
We find that under certain conditions a significant velocity separation is developed between different  ions in the shock downstream and demonstrate that in fast enough shocks ion-ion collisions may trigger fusion and fission events at a relatively high rate.  
Our analysis ignores anomalous coupling through plasma microturbulence, that might reduce the velocity spread downstream 
below the activation energy for nuclear reactions.  A rough estimate of the scale separation in RMS suggests that for shocks propagating in BNS merger ejecta the anomalous coupling length may exceed the radiation length, allowing a considerable composition change behind the shock 
via inelastic collisions of $\alpha$ particles with heavy elements at shock velocities $\beta_u\gtrsim0.25$.    A sufficient abundance of free neutrons in the shock upstream, as expected during the first second after the merger, is also expected to alter the ejecta composition through neutron capture downstream. The resultant change in the composition profile may affect the properties of the early kilonova emission. 
The generation of microturbulence due to velocity separation can also give rise to particle acceleration that might alter the breakout signal in supernovae and other systems. 
\end{abstract}


\section{Introduction}

Radiation mediated shocks (RMS) are inherent in essentially all (strong) stellar explosions  (e.g., various
types of supernovae, low luminosity GRBs, regular GRBs, binary neutron star mergers and tidal disruption events).   
They dictate the properties of the early electromagnetic emission released in the explosion during  the breakout of the shock 
from the opaque envelope enshrouding the source, the detection of which is a primary focus of current and upcoming transient 
surveys (for recent reviews see \citealt{waxman2017,LE2020}).
 The RMS structure and dynamics depend on the progenitor type, the explosion energy and the explosion geometry. In particular, 
the shock velocity at breakout can range from sub-relativistic to mildly relativistic, and in extreme cases even ultra-relativistic.

The gamma-ray flash GRB 170817A that accompanied the gravitational wave signal GW 170817 (for reviews see \citealt{nakar2020,Margutti2021}) is 
a plausible example of a shock breakout signal \citep[e.g.,][]{Kasliwal2017,Gottlieb2018,Beloborodov2020}; the RMS in this source was most likely driven
by the interaction of the relativistic jet expelled by the compact remnant and the merger ejecta \citep{Nakar2018}, but a shock with a much wider opening angle is also a viable possibility \citep{Beloborodov2020}.  While the 
origin of GRB 170817A is still debatable, the presence of a relativistic jet in this system has been confirmed by VLBI observations \citep{Mooley2018}.
Consequently, if this system represents a prototypical BNS merger evolution, the conclusion that a fast shock must cross through at least part of the ejecta at early times seems unavoidable.  
In what follows we demonstrate that the propagation of the RMS through the merger ejecta can significantly alter
 the composition profile of r-process material behind the shock and, potentially,
the kilonova emission if the change in composition affects the opacity and/or the radioactive energy deposition in the ejecta.  

A key issue in RMS theory is how the different plasma constituents are coupled.    
Since the radiation force acts solely on electrons and positrons, it
must be mediated to the ions by some other means. In a sub-relativistic, single-ion RMS this is accomplished through the generation  of an electrostatic field, owing to a tiny charge separation of electrons and ions. However, electrostatic coupling fails in relativistic RMS (RRMS), in which $e^\pm$ pairs are overabundant \citep{Levinson_2020,Arno_plasma_filaments}, and in RMS composed of multi-ion species with different charge-to-mass ratios \citep{derishev2018,le2023}. It has been shown recently \citep{Arno_plasma_filaments} that in an unmagnetized single-ion RRMS, the dominant coupling mechanism is plasma microturbulence, generated by a current filamentation instability driven by the relative drift between the ions and the pairs. The presence of a strong enough (transverse) magnetic field in the upstream flow can give rise to magnetic coupling of pairs and ions which completely suppresses the instability \citep{mahlmann2023}.

The situation can be vastly different in multi-ion RMS.  First, the deceleration rate of ions inside the shock by an electrostatic field depends on the charge-to-mass ratio, 
and since the charge conservation condition in a multi-ion plasma is degenerate a large velocity separation 
between the different ions in the post-deceleration zone is, in principle, allowed (even in sub-relativistic RMS), provided that interspecies friction is ineffective. As shown below, in some cases binary Coulomb collisions effectively couple the ions while in others they are not and detailed calculations are needed to assess whether the relative drift 
established just downstream of the shock surpass the barriers for inelastic nuclear collisions.
Second, since the gyroradius of an ion of mass $Am_p$ is larger by a factor $Am_p/Zm_e$ than that of an electron (or positron), much stronger magnetic fields may be needed to couple all ions. Such strong fields may not be present in most relevant systems. Third, in cases where Coulomb friction is ineffective and cannot suppress plasma instabilities, the coupling of different ions by plasma micro-turbulence would require generation of wave modes different than in a single-ion RRMS.  

In this paper, we construct a semi-analytic, multi-ion RMS model and use it to
compute the structure of a shock propagating in different environments, with emphasis on shocks propagating  in the expanding ejecta of BNS mergers. We find that 
 in {\it Newtonian}  shocks binary Coulomb collisions effectively couple the ions, particularly heavy ions with large atomic numbers. However, the presence of positrons, anticipated 
in fast enough shocks, can lead to the separation of light ions (mainly helium in r-process ejecta) from the bulk of the heavy ions. In case of BNS mergers, We estimate that a shock moving at a velocity 
$\gtrsim0.3c$ (where $c$ is the light speed) in ejecta containing a substantial He abundance, can induce a considerable change in the composition of r-process material behind the shock 
through collisions of alpha particles with heavy ions.
%
%
Moreover, the presence of sufficiently abundant free neutrons in the shock upstream should lead to fission 
through the capture of free neutrons, which do not experience a significant deceleration by the shock, by neutron-rich isotopes downstream of the shock.   
The activation energy for neutron-induced fission ranges from practically zero for $^{235}U$ to a peak of about $40$ MeV 
for elements of mass number $A\approx100$. Consequently, if the shock is driven into the ejecta while free neutrons are abundant (i.e., before nucleosynthesis freezout), 
a considerable impact on the abundance evolution of r-process
elements by neutron capture is anticipated prior to the kilonova emission
in regions where the shock velocity $\gtrsim \, 0.2 c$. 
These gross estimates are supported by detailed calculations presented below.

The plan of the paper is as follows: In Sec. \ref{multi-ion model} we introduce the Multi-ion RMS model. In Sec. \ref{Analytic}
we provide analytic estimates of ion-ion collision energies and collision frequencies.  In Sec. \ref{C-S} we present 
numerical solutions of the shock equations in the absence of interspecies friction for different compositions of the upstream plasma.  In Sec. \ref{friction}
we modify the model to include Coulomb friction forces.  In Sec. \ref{pairs_role} we consider the effect of electron-positron pairs on the shock structure.
In Sec. \ref{sec:free_neutrons} we estimate the fission rate by capture of free neutrons.
In Sec. \ref{sec:BNS_mergers_applc} we discuss the applications
to BNS mergers.  A detailed derivation of the shock equations can be found in appendix \ref{Eq Dev}, followed by the inclusion of coulomb interactions detailed in appendix \ref{coul_fric_eq}.
\section{The multi-ion shock model}\label{multi-ion model}

We construct a semi-analytic model for a Newtonian RMS that propagates in a neutral plasma consisting of electrons and 
a collection of ions having different charge-to-mass ratios.   For simplicity, the shock is assumed to be infinite, planar and in a steady state.
Adopting the approach of  \cite{BPb1981,Blandford_1981}, we employ the diffusion approximation to compute the transfer of radiation through the shock.
A comparison of analytic solutions with full Monte-Carlo simulations \citep{ito2020newt} indicates that the diffusion approximation is quite accurate even at shock 
 velocities $\lesssim 0.3c$ or so, justifying this treatment.  
In contrast to the single-fluid approach invoked in all previous Newtonian RMS models, 
in which all plasma constituents are assumed to be tightly coupled, here we self-consistently compute the electrostatic coupling between 
the electrons and ions, imposed by the charge separation induced by the relative drift between the different multi-fluid components.
This is accomplished by solving the energy and momentum equations of the radiation and the multi-fluid plasma, together with
Maxwell's equations, taking into account the electrostatic force acting on the charged fluids. 
Our treatment is similar to that adopted in \cite{Levinson_2020} for the single-ion relativistic RMS, but with modifications necessary 
for the inclusion of multi-ion species, and with the beaming approximation for the radiative transfer replaced by the diffusion approximation. 

To simplify the analysis, we neglect pair creation and nuclear transmutations, although, as
shown below, fission and fusion might be important in the immediate downstream of fast shocks.  We emphasize that our main goal here is not to 
present detailed calculations  of the composition profile behind the shock, but merely to demonstrate that a substantial change in composition is likely to occur. 
The neglect of pair production is justified for most exploding systems, but not for the extreme densities anticipated 
in BNS merger ejecta (see Sec. \ref{sec:BNS_mergers_applc} for a detailed discussion). 
We tend to believe that inclusion of pair creation, while considerably complicating the solutions, will not 
alter significantly the essence of the results obtained with the simplified model constructed below.  We intend to incorporate pair production into our model in a future publication.

Our analysis also ignores potential generation of plasma waves by ion beam instabilities.
The resultant plasma microturbulence can provide additional coupling mechanism on microscopic scales, that might alter the ion velocity distribution in the 
immediate downstream.  To gain some insight into the effect of such anomalous coupling, we incorporate, in Sec. \ref{friction}, 
a phenomenological interionic friction model into the shock equations (apart from binary Coulomb collisions).   In Sec. \ref{sec:BNS_mergers_applc} we estimate the ratio between the radiation 
and kinetic scales, and discuss the implications for the anomalous coupling length.  Based on this estimate, we speculate that for RMS in BNS mergers, 
anomalous coupling may not alter significantly the results obtained in Sec. \ref{sec:r-process} in the absence of anomalous friction. 
For other exploding systems the effect of plasma instabilities is unclear.

\subsection{Governing equations}\label{model eq}

We solve the shock equations in the rest frame of the shock, in which the flow is steady.  We choose a coordinate system 
such that the upstream flow moves in the positive $\hat{x}$ direction. All fluid quantities are henceforth measured in 
the shock frame, unless otherwise stated.  Let $n_e$ denotes the electron density and $\beta_e$ its velocity (in units of $c$), and 
$n_\alpha$, $\beta_\alpha$, $Z_\alpha$, $A_\alpha$ the density, velocity, atomic number and mass number, respectively, of 
ion species $\alpha$ (here, $\alpha =$ He, Au, etc.).  The charge-to-mass ratio of the $\alpha$ ion is then $Z_\alpha/A_\alpha$.
Since our analysis ignores pair production and nuclear transmutation (although, as explained below, they might be important downstream),
all particle fluxes must be conserved across the shock. In particular, $j_e \equiv n_e\beta_e = n_{e,u}\beta_u$
and $j_\alpha \equiv n_\alpha\beta_\alpha = n_{\alpha,u}\beta_u$, here subscript $u$ labels fluid quantities far upstream of the shock.
For the upstream conditions anticipated (at early times) in BNS merger ejecta all ions are likely to be fully ionized. Hence, $Z_\alpha$ is
a constant for all $\alpha$.  Charge neutrality of the upstream plasma implies $\sum_\alpha Z_\alpha n_{\alpha,u} - n_{e,u}=0$, from which one obtains the relation
\begin{equation}\label{eq:continuity}
    \sum_\alpha Z_\alpha j_\alpha - j_e =0,
\end{equation}
that holds everywhere.

The radiation force inside the shock acts solely on the electrons. This leads to a relative drift between the electrons and the different ion species
which, in turn, induces charge separation, $\rho_e =e(\sum_\alpha Z_\alpha n_\alpha - n_e)\ne 0 $, and the consequent generation of an electrostatic field, $\pmb{E} = E(x)\hat{x}$.  
The change of this electric field across the shock is governed by Gauss' law:
\begin{equation}
 \nabla\cdot\pmb{E} = \frac{dE}{dx} = 4\pi \rho_e. 
\end{equation}
The last equation can be rendered dimensionless by transforming to the coordinate $d\tau = \sigma_T n_e dx$, here $\sigma_T$ is the Thomson cross-section,
and normalizing the electric field $E$ by the fiducial field $E_0 = m_e c^2j_e \sigma_T/e$, viz., $\tilde{E} = E/E_0 = eE/m_e c^2j_e \sigma_T$. This yields
\begin{equation}
 \frac{d \tilde{E}}{d\tau} = \chi \left(\sum_{\alpha}^{} \frac{Z_\alpha j_\alpha \beta_e}{j_e \beta_\alpha} - 1\right),\label{eq:Gauss}
\end{equation}
with 
\begin{equation}\label{eq:chi}
\chi = \frac{4\pi e}{\sigma_T E_0} 
\approx 10^{12} \left(\frac{n_{e,u}}{10^{25} \, \text{cm}^{-3}}\right)^{-1}\beta_u^{-1}.
\end{equation}
The density is normalized here to the fiducial density anticipated in BNS merger ejecta about a second after the merger (Eq. \ref{eq:density_BNS}).  In SNe and GRBs 
the anticipated density is much smaller, $n_{e,u}\lesssim 10^{15} \, \text{cm}^{-3}$. 

The energy and momentum equations of the multi-fluid system are derived in appendix \ref{Eq Dev} under the assumption that the electron and ion fluids are cold.
This simplifying assumption is justified by the fact that the pressure inside the shock is completely dominated by the diffusing radiation.  
In the dimensionless form, the reduced set of equations read:

\begin{align}
 &   \frac{d}{d\tau}(\beta_e + \pi_\gamma)  = \text{-\~{E}}, \label{eq:1}\\
  &   \frac{d}{d\tau}\beta_\alpha = \mu \frac{Z_\alpha   \beta_e}{A_\alpha \beta_\alpha } \text{\~{E}},\label{eq:2}\\
    &   \frac{d}{d\tau}\pi_\gamma= \frac{1}{\beta_e}\frac{d}{d\tau}(4\pi_\gamma\beta_e-\frac{d}{d\tau}{\pi_\gamma}),\label{eq:4}
\end{align}
where $\pi_\gamma$ is the normalized radiation pressure, $\pi_\gamma = p_\gamma/m_ec^2 j_e$, and $\mu = m_e/m_p$ is the electron-to-proton mass ratio.
Equations (\ref{eq:Gauss})-(\ref{eq:4}) form a closed set that can be solved once the values of $\beta_u$, $j_e$, $j_\alpha$ and $Z_\alpha/A_\alpha$ are specified.
The far upstream value of the electric field is $\tilde{E}(x\to -\infty)=0$.

\subsection{Numerical integration} \label{numeric}

Solutions for the multi-fluid RMS structure of a shock that propagates in photon-poor upstream are obtained in section \ref{C-S} by numerically solving 
Eqs. \eqref{eq:Gauss}-\eqref{eq:4}, assuming a cold, neutral plasma in the far upstream.
We note that the shock equations, Eqs. \eqref{eq:Gauss}-\eqref{eq:4}, are invariant under translations,
hence, the location of the origin can be chosen arbitrarily. 
For convenience, we start the integration in all runs at $\tau=0$, with $\beta_{e,u}=\beta_{\alpha,u}=\beta_u$, $\pi_{\gamma,u} = 0.001$,
$\tilde{E} =0$, and $\tilde{T}^{0x}_{\gamma,u} = 0$ as initial values (at $\tau=0$), 
where $\tilde{T}^{0x}$ is the radiation energy flux, defined in Eq. \ref{eq:radiation_E_flux}. 
With this choice the radiation force at $\tau=0$ is given by 
\begin{equation}
 \frac{ d\pi_{\gamma,u}}{d\tau} = 4\pi_{\gamma,u}\beta_u = 0.004\beta_u.\label{eq:rad_force_bd}
\end{equation}
We verified that as long as
$\pi_{\gamma,u}$ is small enough (i.e., the upstream is photon poor), the solution converges, except for a shift in the location of the shock which, as explained above, is arbitrary.

Since only the ratio $j_\alpha/j_e=n_{\alpha,u}/n_{e,u}$ appears in the above equations, we find it convenient to 
normalize the upstream densities to the electron density $n_{e,u}$, and define $\tilde{j}_\alpha=j_\alpha/j_e$.
To save computing time, we use $\chi=10^6$ instead the more realistic value given in Eq. (\ref{eq:chi}).  We checked that the
solution is highly insensitive to the value of $\chi$, provided it is not too small ($> 10^3$ roughly).

\section{Estimation of ion-ion collision energy and collision frequency} 
\label{Analytic}

A qualitative description of the multi-flow dynamics inside the shock is as follows:
At the outset of the shock transition layer, the incoming electron fluid is decelerated and thus compressed, by the radiation force 
(the gradient of the radiation pressure in the diffusion limit).  This creates charge separation which, in turn, induces an electrostatic field
inside the shock. This electric field tends to decelerate the ions and accelerate the electrons, thereby providing coupling of the various plasma components.
Charge neutrality is regained once the radiation pressure saturates (and the radiation force vanishes), completely suppressing the electric field.  
In the absence of any other coupling mechanism (e.g., binary collisions, plasma turbulence; see section \ref{friction} for further discussion), 
relative drifts between ions with different charge-to-mass
ratio ensues in the post-deceleration zone (immediate downstream), depending on the value of $Z_\alpha/A_\alpha$. The slowest ion in the immediate 
downstream, which we henceforth term 'agile', is the one with the largest $Z/A$ ratio among all ions present in the upstream flow.

While the full shock structure can only be found upon numerical integration of Eqs. \eqref{eq:Gauss}-\eqref{eq:4}, an expression for the immediate downstream velocity
of each ion in terms of the electric potential across the shock can be obtained analytically by integrating Eq. \eqref{eq:2} alone.  This enables reasonably accurate 
estimates of the relative drift between the ions in the absence of effective friction and the consequent collision energies.  In what follows, we provide estimates of the relative drifts, collision energies, and collision frequencies of ions in the immediate downstream when inter-species friction is ignored.

\subsection{Ion velocity separation}\label{sep} 

Since ions do not experience the radiation force in our model, the change in their energy across the shock must
equal the work done by the electrostatic field.   When normalized to the upstream baryon energy, $m_pc^2\beta_u^2/2$, the latter reads:

\begin{equation}\label{eq:WE}
w_E \equiv \int_{-\infty}^{\infty}\frac{2E\cdot e}{m_p c^2 \beta_{u}^2}dx.
\end{equation}

Integrating Eq. \eqref{eq:2}, an expression for the downstream terminal velocity of ion species $\alpha$ is obtained in terms of $w_E$:
\begin{equation}\label{eq:7}
    \beta_{\alpha,d}=\beta_u \sqrt{1+w_E\frac{Z_\alpha}{A_\alpha}}.
\end{equation}

The velocity difference between ion species $\alpha$ and $\alpha^\prime$ now reads:
\begin{equation}\label{eq:8}
    \Delta\beta_{\alpha,\alpha^\prime} = \beta_u \biggl[\sqrt{1+w_E\frac{Z_{\alpha}}{A_{\alpha}}}-\sqrt{1+w_E\frac{Z_{\alpha^\prime}}{A_{\alpha^\prime}}}\biggr]. 
\end{equation}
This relation is exact, but the value of $w_E$ is unknown.  To estimate $w_E$ we note that 
for the agile ion, the one with the largest charge to mass ratio, $\beta_{agile,d}^2<<\beta_{u}^2$. Adopting $\beta_{agile,d}=0$ in Eq. \eqref{eq:7} gives
$w_E\approx-\frac{A_{agile}}{Z_{agile}}$.  Substituting the latter expression into Eq. \eqref{eq:8}, we can evaluate $\Delta\beta_{\alpha,\alpha^\prime}$
for all ions once the ratio $A_{agile}/Z_{agile}$ is specified. 

As an example, consider a mixture of hydrogen and helium only.  The agile ion here is $H$, for which $A_H/Z_H=1$ and
 $w_E\approx- A_H/Z_H=-1$, giving $\Delta\beta_{He,H}\approx 0.21 \bigl(\beta_u/0.3\bigr)$.
 This result is in good agreement with the numerical solution derived in section \ref{C-S} (see Fig. \ref{fig:H_He}), for which
 $\Delta\beta_{He,H}\approx0.173$, for a shock velocity of $\beta_u=0.3$.
 In the case of heavy r-process composition (section \ref{sec:r-process}), we find $w_E\approx-A_{He}/Z_{He}\approx-2$ for the agile ion $He$. 
 This yields, for example, velocity separation of $\Delta\beta_{Ga,He}\approx0.095 \bigl(\beta_u/0.3\bigr)$ 
between $Ga$ (with $Z_{Ga}=31$ and $A_{Ga}=69$) and $He$, again in good agreement with the numerical solution ($\Delta\beta_{Ga,He}\approx0.091$) presented in Fig. \ref{fig:r_process}.
We find good agreements in all other cases tested and conclude that Eq. \eqref{eq:8} can be safely used to calculate relative drifts between ions.

The above derivation suggests that the downstream conditions are highly insensitive to the upstream density of the agile ion.  As will be shown in Sec. \ref{C-S}, this 
may be of utmost importance for astrophysical applications. To see this, note first that Eq. \eqref{eq:7} yields the exact result 
$w_E = -(A_{agile}/Z_{agile})[1-(\beta_{agile,d}/\beta_u)^2]$, which depends on $\beta_{agile,d}$ to second order.  From Eq. \eqref{eq:7} we 
also deduce that the downstream velocity of the non-agile ions, for which $\beta_{\alpha,d} \gg \beta_{agile,d}$, depends on $\beta_{agile,d}$ to the same
order, and since the flux $j_\alpha$ is conserved across the shock, the downstream density, $n_{\alpha,d}= j_\alpha/\beta_{\alpha,d}$, also depends on $\beta_{agile,d}$ to second order.  
From Eqs. \eqref{eq:1} and \eqref{eq:4} we find, upon neglecting 
the electron contribution to the energy budget:
$$m_pc^2\beta_u^2 w_E = 8 e \int_{-\infty}^{\infty} E(x) \frac{\beta_{e,d}}{\beta_e(x)}dx.$$
Since to order $O(\beta_{agile,d}/\beta_u)^2$ the left hand side of this equation is constant, it implies that $\beta_e(x)$ and, hence, $\beta_{e,d}$ and $n_{e,d}$,
must be preserved to the same order.   The charge neutrality condition downstream,
 $Z_{agile}n_{agile,d} = n_{e,d} -\sum_{\alpha\ne agile} Z_{\alpha,d} n_{\alpha,d}$,
then implies that the downstream density of the agile ion, $n_{agile,d}$, must also be preserved to this order and is, therefore,
highly insensitive to the upstream density $n_{agile,u}$. In other words, the compression ratio of the agile ion increases as its mass fraction upstream decreases
in order to maintain $n_{agile,d}$ at the level required for charge neutrality downstream.  This argument applies 
 provided the diffusion approximation remains valid.

\subsection{Collision energy}\label{collision}

In order to compute the collision energy in $\alpha - \alpha^\prime$ collisions,
it is convenient to transform to the center of momentum frame (COM).  One then finds 
\begin{equation}\label{eq:9}
    E_{COM} = \frac{1}{2}\frac{A_{\alpha}A_{\alpha'}}{A_{\alpha}+A_{\alpha'}} m_p c^2 \Delta\beta_{\alpha,\alpha'}^2,
\end{equation}
where $\Delta\beta_{\alpha,\alpha'}$ is given by Eq. (\ref{eq:8}).  As an example, consider the $H-He$ collision.  With $A_H=1, A_{He}=4$,  Eq. \eqref{eq:8} 
yields $\Delta \beta_{H,He}=0.17 (\beta_u/0.3)$, and from Eq. \eqref{eq:9} we have $E_{COM} = 11.5 (\beta_u/0.3)^2$ MeV.

In general, we find that for light elements $E_{COM}\sim (1-20) (\beta_u/0.3)^2$ MeV and for heavy elements  $E_{COM}\sim (10 - 300) (\beta_u/0.3)^2$ MeV.
Thus, at shock velocities $\beta_u \gtrsim 0.5$, the collision energy can exceed  $\sim $ 1 GeV for certain combinations of ions. 

Coulomb barriers range between about 1 and 300 MeV for most elements. 
Therefore, ion-ion collisions behind a fast enough shock may conceivably cause significant composition change.

\subsection{Collision rate}

The geometrical cross-section for inelastic collisions of ion species $\alpha$ and $\alpha^\prime$ is given approximately by 
\begin{equation}
\sigma_{\alpha,\alpha'} = \pi[(A_{\alpha}^{1/3}+A_{\alpha'}^{1/3}){r_p}]^2 \approx 0.068\sigma_T(A_{\alpha}^{1/3}+A_{\alpha'}^{1/3})^2, 
\end{equation}
where $r_p\sim 0.43\, r_e \sim1.2fm$, with $r_e=\frac{e^2}{m_e c^2}$ being 
the classical electron radius, and $\sigma_T = \frac{8\pi}{3}r_e^2$ the Thomson cross section \citep[e.g.,][]{BASS1973139,Kox_sigma}.

The corresponding collision frequency is given by 
\begin{equation}\label{eq:nu_pp}
\nu_{\alpha\alpha^\prime} = n_{\alpha',d} \sigma_{\alpha,\alpha'}c \lvert \Delta\beta_{\alpha,\alpha'} \lvert.
\end{equation}
To obtain the total collision frequency, we integrate over all $\alpha-\alpha'$ interactions with minimal energy $E_{min}$. This corresponds to a region in the shock with $\Delta\beta_{\alpha,\alpha'}>\sqrt{\frac{2(A_\alpha+A_{\alpha'})E_{min}}{A_\alpha A_{\alpha'}m_p c^2}}$. We use the coulomb barrier between any two nuclei as the minimal value. $E_{min}$ is obtained by Eq. (1) in \cite{royer2021}.

The mean number of interactions, within the relevant region $\Delta\tau$, for ion of type $\alpha$ upon ion of type $\alpha'$ is:
\begin{equation}\label{eq:10}
\begin{split}
    \Delta\tau_{\alpha,\alpha'}&=\int_{\Delta\tau}\frac{\nu_{\alpha\alpha^\prime} l_T}{\beta_{\alpha,d}\cdot c}d\tau
    =\int_{\Delta\tau} \frac{n_{\alpha',d} 0.068\sigma_T (A_{\alpha }^\frac{1}{3}+A_{\alpha'}^\frac{1}{3})^2}{n_{e} \sigma_T } \cdot \frac{\lvert \Delta\beta_{\alpha,\alpha'} \lvert}{\beta_{\alpha,d}}d\tau\\ 
    &= 0.068\int_{\Delta\tau} \Tilde{j}_{\alpha'}\frac{\beta_{e}}{\beta_{\alpha',d}} \cdot\frac{\lvert \Delta\beta_{\alpha,\alpha'} \lvert}{\beta_{\alpha,d}} (A_{\alpha }^\frac{1}{3}+A_{\alpha'}^\frac{1}{3})^2d\tau\\
    &= 0.068 \cdot(A_{\alpha }^\frac{1}{3}+A_{\alpha'}^\frac{1}{3})^2  \Tilde{j}_{\alpha'}\int_{\Delta\tau}\beta_{e} \bigg| \frac{1}{\beta_{\alpha',d}}-\frac{1}{\beta_{\alpha,d}} \bigg| d\tau.
    \end{split}
\end{equation} 
We used $l_T = (\sigma_T n_e)^{-1}$. 

\section{Numerical results}\label{C-S} 
In this section, we present the solution of the shock equations, Eqs. \eqref{eq:Gauss}-\eqref{eq:4}.
To illustrate the basic properties of the solution we begin with a simple example of a pure hydrogen-helium ($H-He$) mixture.  We then 
present solutions for solar composition and for r-process elemental abundances anticipated in BNS merger ejecta.
In all of these examples, the shock velocity is taken to be $\beta_u=0.3$.
The same case studies will be employed in subsequent sections, where friction (Sec. \ref{friction}) and pairs (Sec. \ref{pairs_role}) will be 
included in the model, changing the shock structure in different regimes.

\subsection{$H-He$ mixture}\label{sec:H-He}
\begin{figure}
\centering
    \includegraphics[width=1\columnwidth]{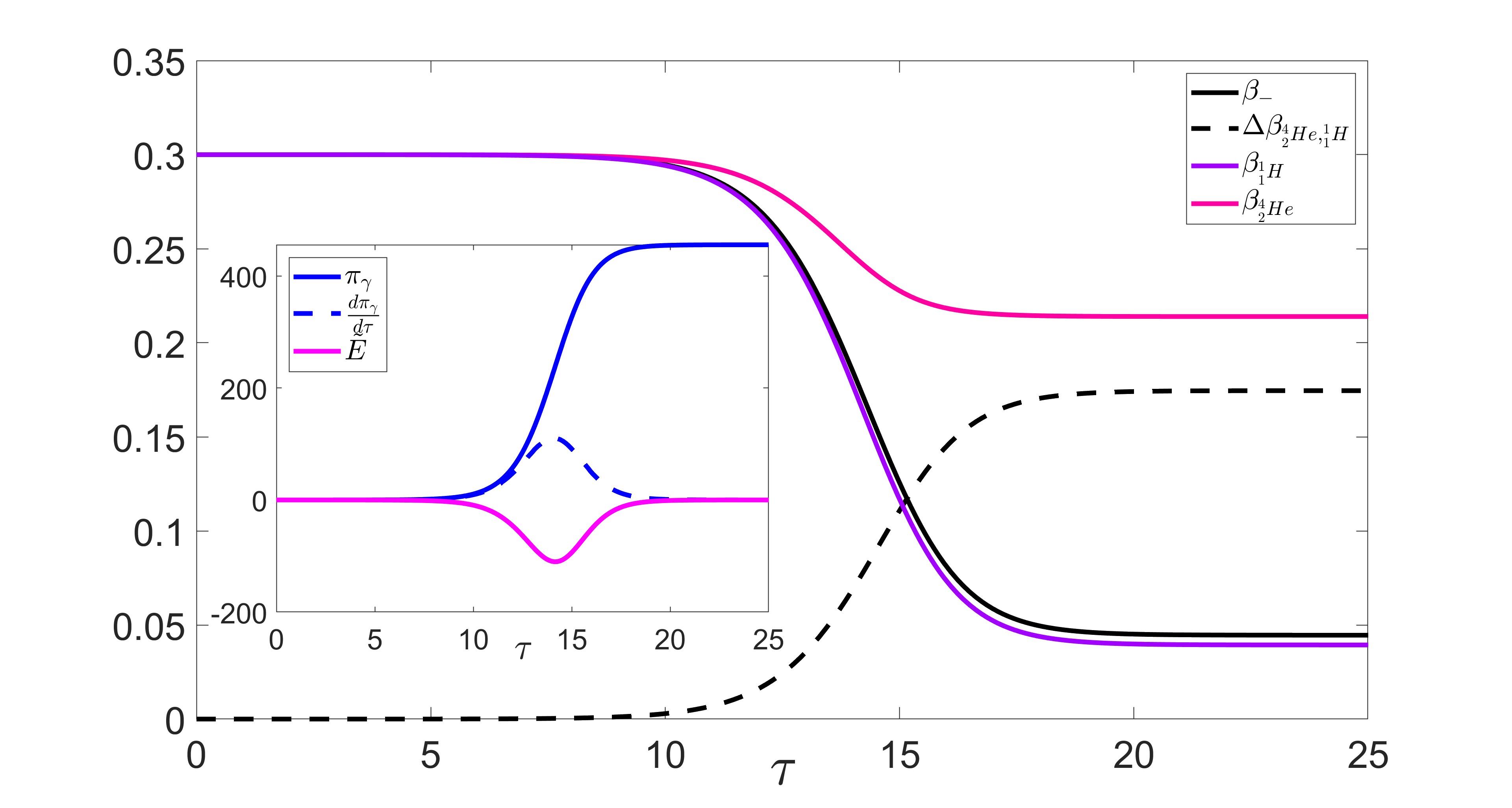}
    \caption{Shock structure solution obtained for a pure hydrogen-helium mixture. 
    The solid blue and red lines represent the velocity profiles of $H$ and $He$, respectively, as indicated, and the solid black line the velocity profile of the electrons.
    The black dashed line delineates the relative drift velocity $\Delta \beta_{He,H}=\beta_{He} - \beta_H$.  The inset shows the electric field (red line), radiation force 
    (blue dashed line) and radiation pressure (blue solid line).
    As seen, the relative drift velocity approaches $\Delta \beta_{He,H}=0.173$ in the post-deceleration zone, at $\tau=17$ roughly, where the electric field and radiation 
    force vanish.}
    \label{fig:H_He}
\end{figure}

In our first example, we compute the structure of a shock propagating at a velocity $\beta_u=0.3$ 
in a medium composed of $H-He$ mixture, with  relative abundances of $X=0.75$, $Y=0.25$.
With our normalization, this translates to the far upstream values $\Tilde{j}_{H,u}=\frac{6}{7}$ and $\Tilde{j}_{He,u}=\frac{1}{14}$ of $H$ and $He$ fluxes, respectively.
The resultant shock structure is depicted in Fig \ref{fig:H_He}.  
As seen, the small radiation force invoked at $\tau=0$, Eq. \eqref{eq:rad_force_bd},
leads initially to the deceleration of the electrons, and the consequent generation of a negative electric field (magenta line in the inset)
by the resultant charge separation. 
This electric field, in turn, decelerates the ions (Eq. \ref{eq:2}), and counteracts the radiation force acting on the electrons (Eq. \ref{eq:1}).  
As a result, the agile $H$ ions remain tightly coupled to the electrons, owing to their larger charge-to-mass ratio, whereas 
the $He$ ions quickly decouple.  Charge neutrality is eventually regained (at $\tau\approx 17$), whereupon the net force acting on the system (the sum of electric and 
radiation forces) vanishes, and the plasma continues to stream undisturbed.  The relative drift velocity (marked by the dotted black line) approaches $\Delta\beta_{H,He}=0.173$ in the post deceleration zone, 
and the corresponding collision energy is $\sim12.16$ MeV.

\subsection{Solar abundances}
\begin{figure}
\centering
\subfloat[$X_H=0.75$]{
	   \centering
	   \includegraphics[width=1\columnwidth]{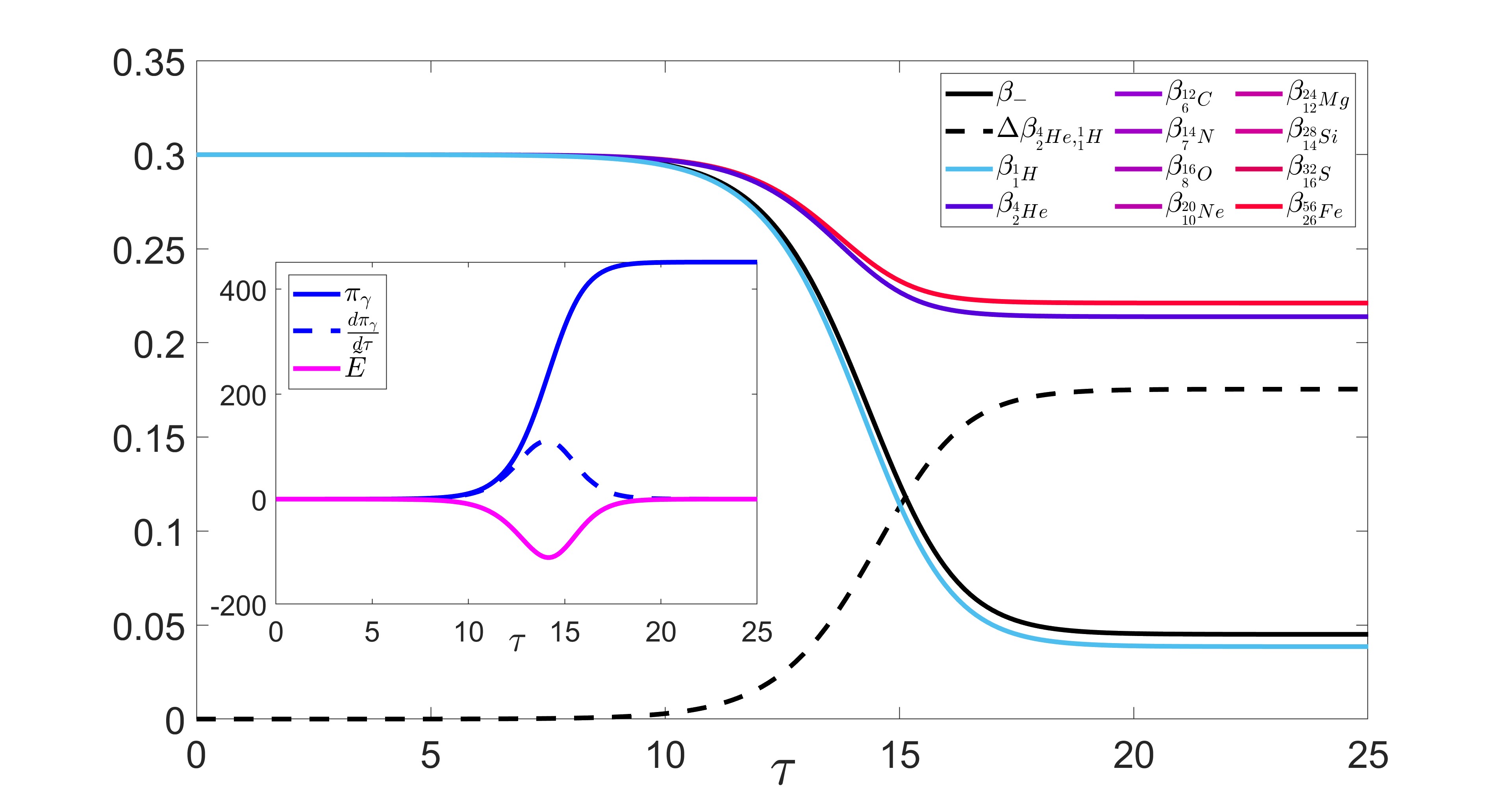}
	\label{fig:solar}}
 \hfill
 \subfloat[$X_H=0.01$]{
	   \centering
	\includegraphics[width=1\columnwidth]{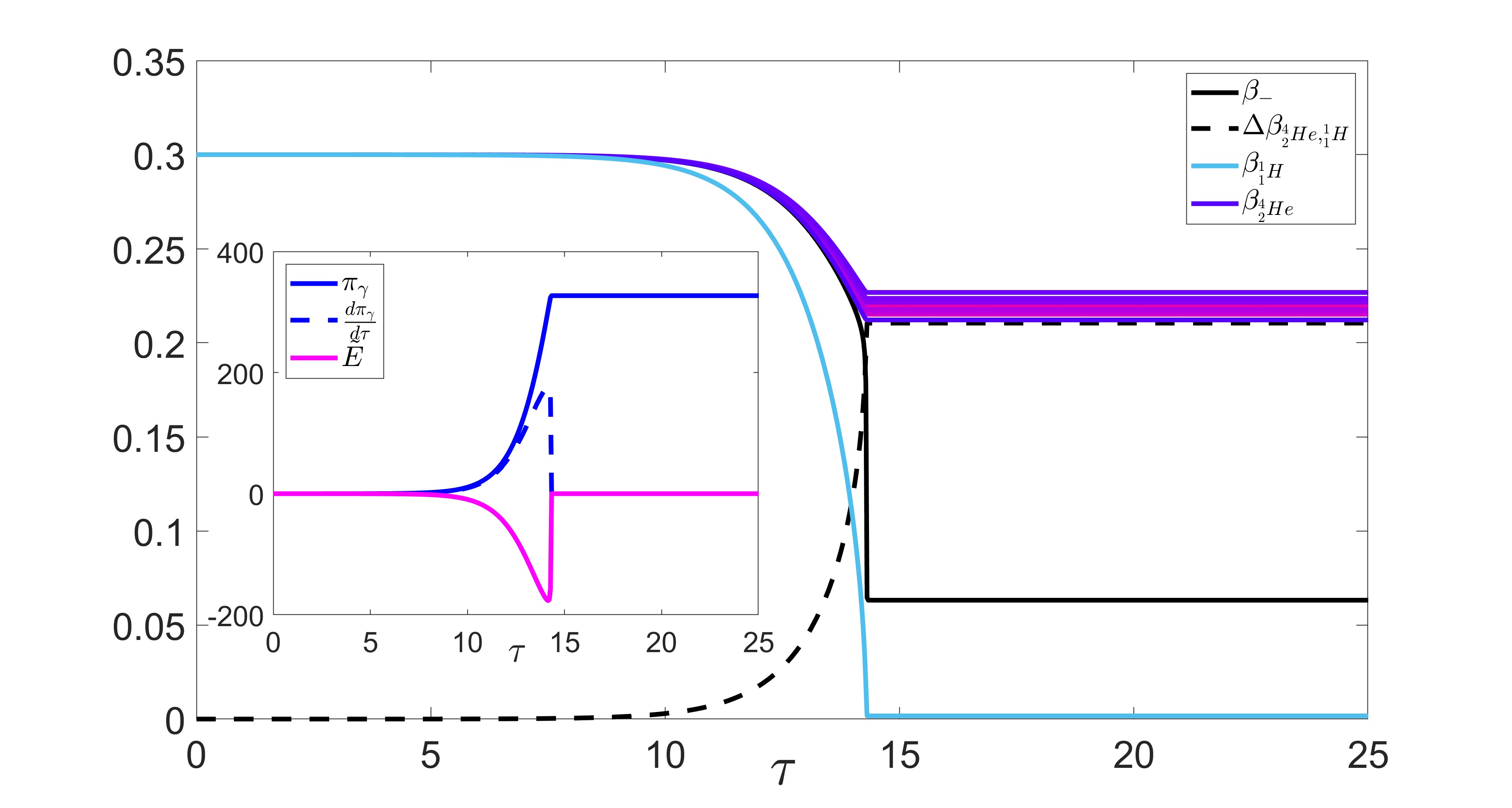}
	\label{fig:solar_low_H}}
\hfill
 
\caption{Solutions for a shock with solar composition for different hydrogen mass fraction ($X_H$).
The large difference in $Z/A$ between the hydrogen and all other ions creates a clear distinction between the $H$ fluid and the cluster of all other ions.   
}
    \label{fig:full solar}
\end{figure}

In our second case, we use the solar elemental abundances given in \citet{lodders19} (see Table 4 there). 
Figure \ref{fig:solar} shows the shock profile.  
In similarity to the previous example, the hydrogen fluid remains tightly coupled to the electron fluid throughout the shock, but
all other ions decouple and form a separate component, with a velocity spread considerably smaller than the relative drift between the $H$ ions
and all other ions. The collision energies range between practically zero and $20$ MeV, for the shock velocity adopted, 
and are dominated by collisions of hydrogen with heavy ions.  Above roughly $5-10$ MeV, proton capture reactions for many elements (e.g. C, N and O) are quite 
rapid, with cross sections in excess of 100 mb (\cite{p_capture}). We thus anticipate such reactions to ensue at shock velocities $\beta_u\gtrsim0.2$,
provided that the relative velocity drifts are maintained for distances comparable to the shock width. 
While there is no detectable signature of H in the spectra of type Ib/c SNe, the progenitors of these SNe might nonetheless contain a small fraction of H (\citealt{Hachinger2012}). In fact, a study of the confirmed single type Ib progenitor that was detected in pre-explosion images (SN iPTF13bvn), suggests that H constitutes almost 1\% of the progenitor's mass \citep{Gilkis2022}. Most of the H mass resides in the outer part of the envelope, where its fraction can reach tens of percents. 
We therefore considered also cases with relative H abundance substantially smaller than solar, specifically, in the range $10^{-5} - 1$ solar (Figure \ref{fig:solar_low_H}).
We find that over this range the ratio between the downstream densities of H and the remaining elements is practically
independent of the mass fraction of H ahead of the shock. As explained at the end of Sec. \ref{sep}, this is a consequence 
of the charge neutrality condition and the fact that the agile
ion is most affected by the electrostatic field inside the shock.
 
In type IIb SNe, velocities of $0.1\sim0.15c$ may be obtained at the ejecta front \citep{Chevalier_2010}, which is marginal for 
proton capture. However, considerably higher shock velocities are conceivable in Type Ib/c, and if indeed contains H ions, proton 
capture reactions might be important in those systems, and can alter the composition in regions where the shock is fast enough. 


\subsection{r-process composition in BNS mergers}\label{sec:r-process}
As a final (and most interesting) example we consider a shock propagating in BNS merger ejecta consisting of r-process material. The composition of various fluid elements in the ejecta of BNS merger depends on its initial condition, mostly the initial electron-fraction and its expansion velocity. However, all fluid elements have in common a dominant fraction of r-process material, free neutrons (which we include later in Sec. \ref{sec:free_neutrons}), and a non-negligible fraction of He. Since He is the most agile element expected in the ejecta, its presence has a significant effect on the velocity spread. Typical He mass fractions found in various simulations range between $10^{-4}$ and $0.1$ (\citealt{Tarumi2023} and references therein). We calculated the velocity spread for far-upstream compositions with these two extreme values, whereas for the r-process elements we adopt 
the abundances given by \cite{Goriely_r_process}.   

\begin{figure}
\centering
\includegraphics[width=1\columnwidth]{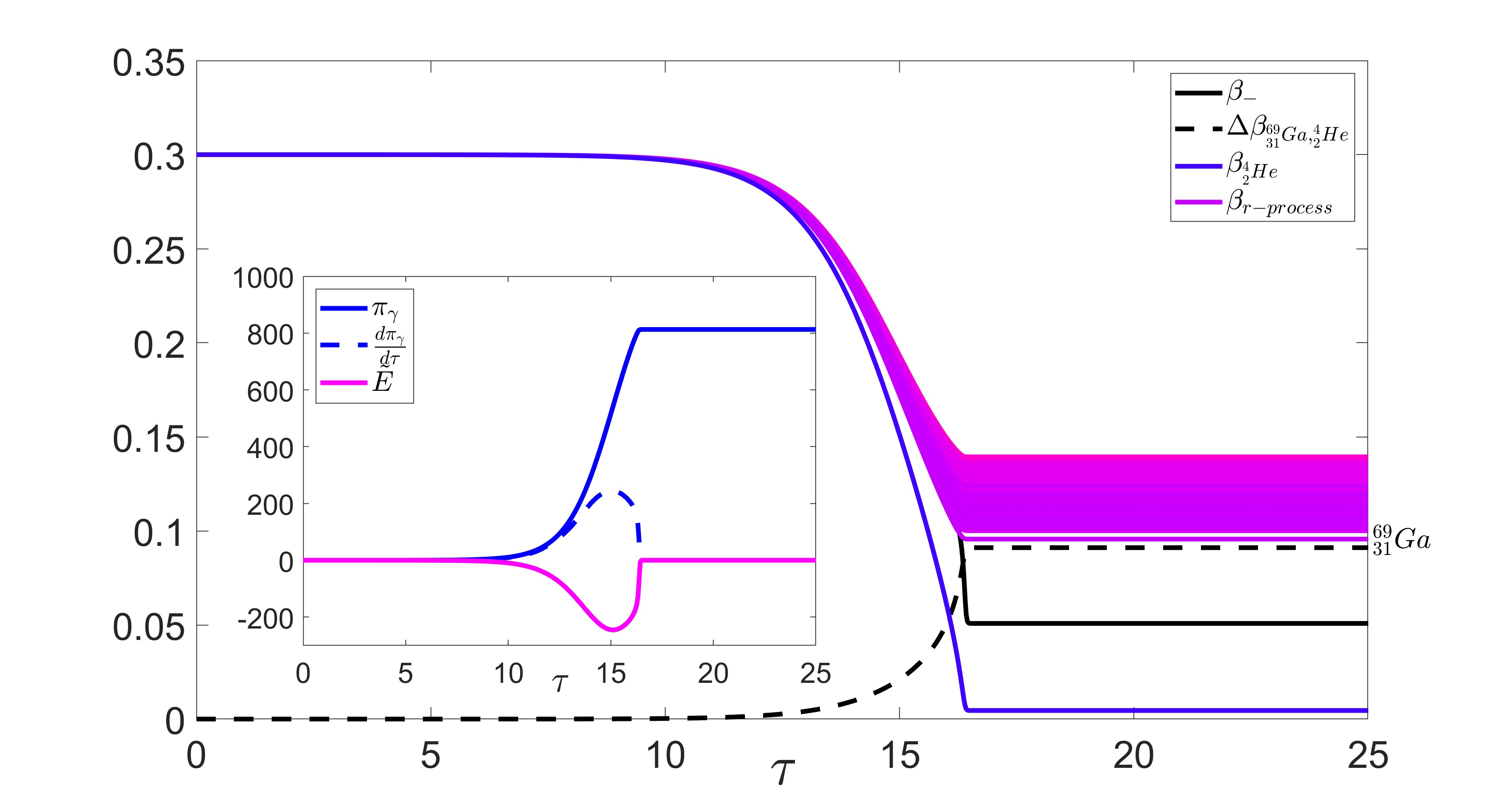}
\caption{Same as Fig. \ref{fig:full solar} for a shock propagating in a BNS merger ejecta, containing He ions with 
a mass fraction of $X_{H_e}=10^{-2}$. The $_{31}^{69}Ga$ ions represent the r-process span, having the lowest downstream velocity.}
    \label{fig:r_process}
\end{figure}

The results are summarized in Fig. \ref{fig:r_process}, where we use the mass number $A$ instead of $Z$ to label the different ions, since
different isotopes of the same elements are present in the system.  The He mass fraction in this example is $X_{He}=10^{-2}$. 
As in the case of solar abundance with varying $X_H$, we find that in the range $10^{-4} < X_{He} < 0.1$ that we explored, the ratio between the density of He and the remaining elements in the immediate 
downstream is independent of the mass fraction of He ahead of the shock. 
Nearly all elements would experience collisions with He 
with COM energies well above the Coulomb barrier. This should give rise to nuclear transmutations of many isotopes at shock velocities $\beta_u\gtrsim0.2$,

\section{Inclusion of interionic friction}\label{friction}
Momentum exchange between the different plasma beams can transfer the radiation force to the ions, ultimately 
leading to coupling of all plasma constituents. Binary Coulomb collisions constitute one potential mechanism for such 
interspecies friction. In situations in which this mechanism fails to strongly couple the ions preventing
large relative drifts, it is expected that
the velocity distribution of the multi-ion plasma established in the post-deceleration zone (Figs \ref{fig:H_He}-\ref{fig:r_process}) will
 be prone to various beam instabilities.  This would result in the generation of plasma microturbulence
that can transfer momentum between different ion beams via ion scattering off plasma waves, thereby driving the downstream plasma
towards equipartition.  The anomalous coupling length is expected to depend on the threshold drift velocity for the onset of 
the instability, on the saturation level of the turbulence, and conceivably other factors. 
Moreover, the continuous decrease of the relative drift velocity between different ion species will induce charge separation 
that will affect the electrostatic field.  We note that conversion of the dissipation energy to radiation can only occur over the radiation length scale.
Consequently, if the anomalous coupling length is much smaller than the Thomson length, dissipation via anomalous friction should
first lead to ion heating before the energy will be converted to radiation. (On a global scale this would appear as formation of collisionless subshocks.)  
Furthermore, if the growth of the instability occurs well after decoupling, the ion temperature should be 
high enough to allow inelastic ion-ion collisions by random motions. 
On the other hand, the electrostatic field induced by the interionic friction 
may prevent the early decoupling of ions with small $Z/A$ seen in the solutions presented in Sec. \ref{C-S}, thereby suppressing the rate 
of inelastic ion-ion collisions.  What would actually happen depends, ultimately, on the threshold velocity of the instability and the scale 
separation. In Sec. \ref{sec:BNS_mergers_applc} we provide a rough estimate of the scale separation in RMS, and conclude 
that in BNS mergers the anomalous coupling 
length is likely to be comparable to, or even larger than, the radiation length, whereas in other systems it is most likely much smaller. 
Thus, the situation might be different in different types of systems. 

To gain insight into the effect of interionic friction, we constructed, in Sec. \ref{fric model}, a (simplified) phenomenological friction model
meant to represent unspecified anomalous coupling mechanism. 
The model ignores any threshold effects, and doesn't take into account ion heating, thus, it is rather limited.  Nonetheless, 
it elucidates some basic features that can guide future analysis.  In Sec. \ref{binary collisions} and appendix \ref{coul_fric_eq} we derive friction coefficients for 
binary Coulomb collisions, and present shock solutions that incorporate Coulomb friction between all species.

\subsection{A phenomenological friction model}\label{fric model}

The full derivation of the shock equations with interspecies friction is given in appendix \ref{app_firc_derivation}.  Here 
we provide a succinct account of the friction model.
We suppose that the change in momentum of ion $\alpha$ due to momentum exchange with ion beam $\alpha'$ is
proportional to the relative velocity between ion species $\alpha$ and $\alpha'$, and the density of the $\alpha'$ beam: $\mathcal{F}_{\alpha\alpha'}=-g_{\alpha\alpha'} n_{\alpha'} (\beta_\alpha-\beta_{\alpha'})$, where $g_{\alpha\alpha'}$ is a characteristic friction coefficient.  Momentum conservation requires the symmetry 
$g_{\alpha \alpha'} = g_{\alpha'\alpha}$ (see \ref{app_firc_derivation} for details).
The total friction force exerted on ion $\alpha$ is the sum over all ion beams. In
 terms of the conserved fluxes $\tilde{j}_{\alpha'}$ and the coordinate $\tau$, the modified Eq. \eqref{eq:2} reads:

\begin{equation}\label{eq:11}
    \frac{d}{d\tau} \beta_{\alpha} = \frac{\beta_e}{\beta_{\alpha} A_{\alpha}} \left[ \mu \text{\~{E}} Z_{\alpha} 
    - \sum_{\alpha'}\text{\~g}_{\alpha\alpha'} \frac{\Tilde{j}_{\alpha'}}{\beta_{\alpha'}}(\beta_{\alpha}-\beta_{\alpha'}) \right], 
\end{equation}
here $\text{\~g}_{\alpha\alpha'} \equiv \frac{g_{\alpha,\alpha'}}{m_p \sigma_T}$ defines a dimensionless friction coefficient. 
In what follows we assume for simplicity that all the coefficients are equal:  $\tilde{g}_{\alpha\alpha'} = \tilde{g}$.
This, of course, may not be realistic in practice (and indeed not the case for, e.g., Coulomb collisions; see Sec. \ref{binary collisions}).  
Nonetheless, it allows us to estimate the strength of the friction force required to tightly couple all ions. 

As stated above, momentum conservation is automatically satisfied if the coefficients $g_{\alpha\alpha'}$ are symmetric.  
However, since friction forces are dissipative, additional physics is
needed to specify how the dissipation energy is distributed.   Here, we assume that energy is transferred from ions to electrons 
over a length scale sufficiently short to keep the electron and  ions  fluids cold (i.e., highly supersonic) 
at all times.  Under this assumption the energy equation reduces to Eq. \eqref{eq:app_energ_fric} in appendix \ref{app_firc_derivation}.

%
\subsection{Shock solutions}
\begin{figure}
    \centering
    \includegraphics[width=1\columnwidth]{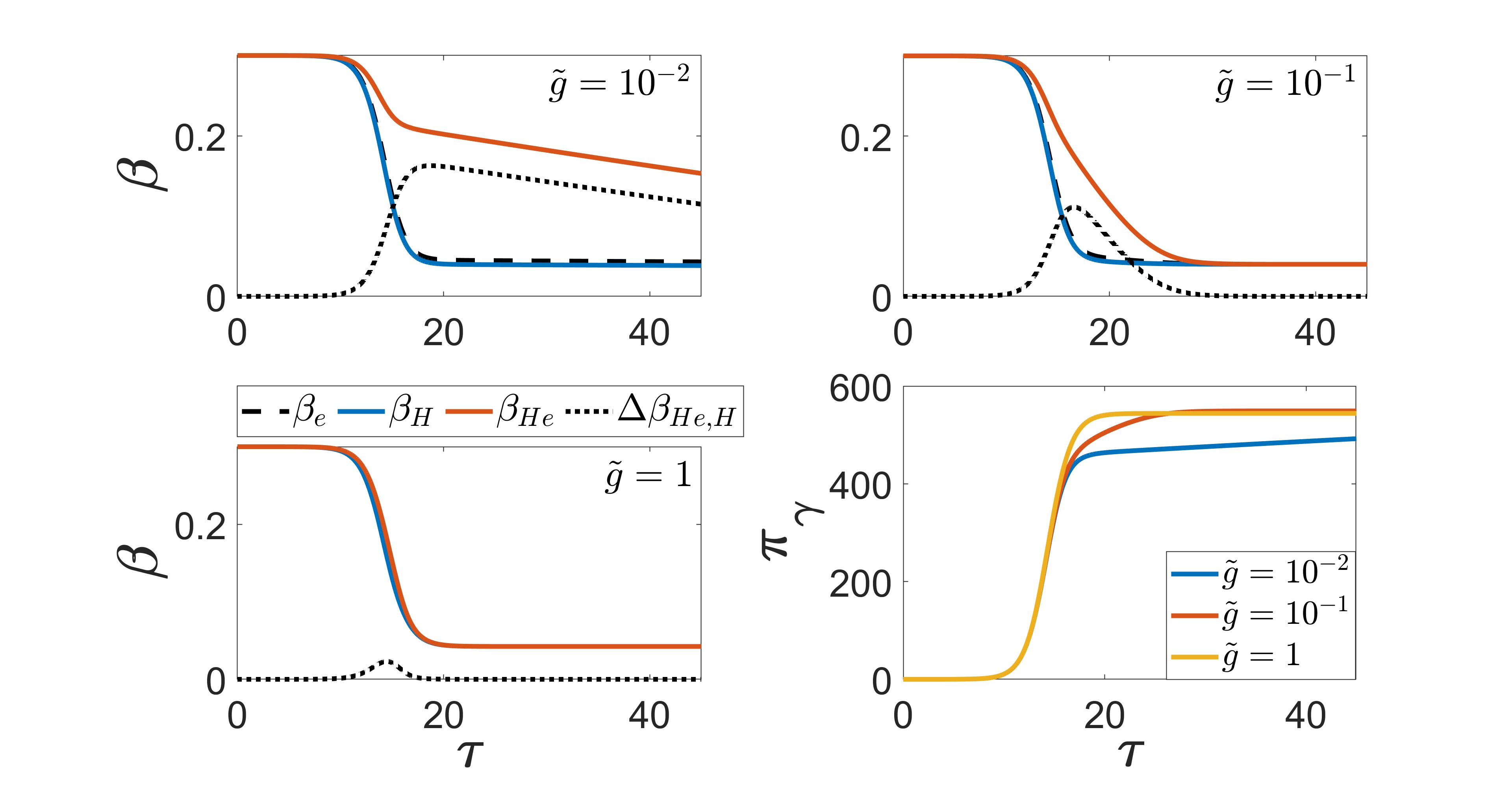}
    \caption{Shock solutions for a pure $H-He$ composition, as in Fig. \ref{fig:H_He}, and different values of the friction coefficient $\tilde{g}$, as indicated.
    The lower-right panel displays the radiation pressure profile for the three cases presented in the other panels ($\tilde{g}= 10^{-2}, 10^{-1}, 1$).  As seen, 
    full coupling occurs for $\tilde{g}\ge1$.}
    \label{fig:H_He_friction}
\end{figure}

\begin{figure}
    \centering
    \includegraphics[width=1\columnwidth]{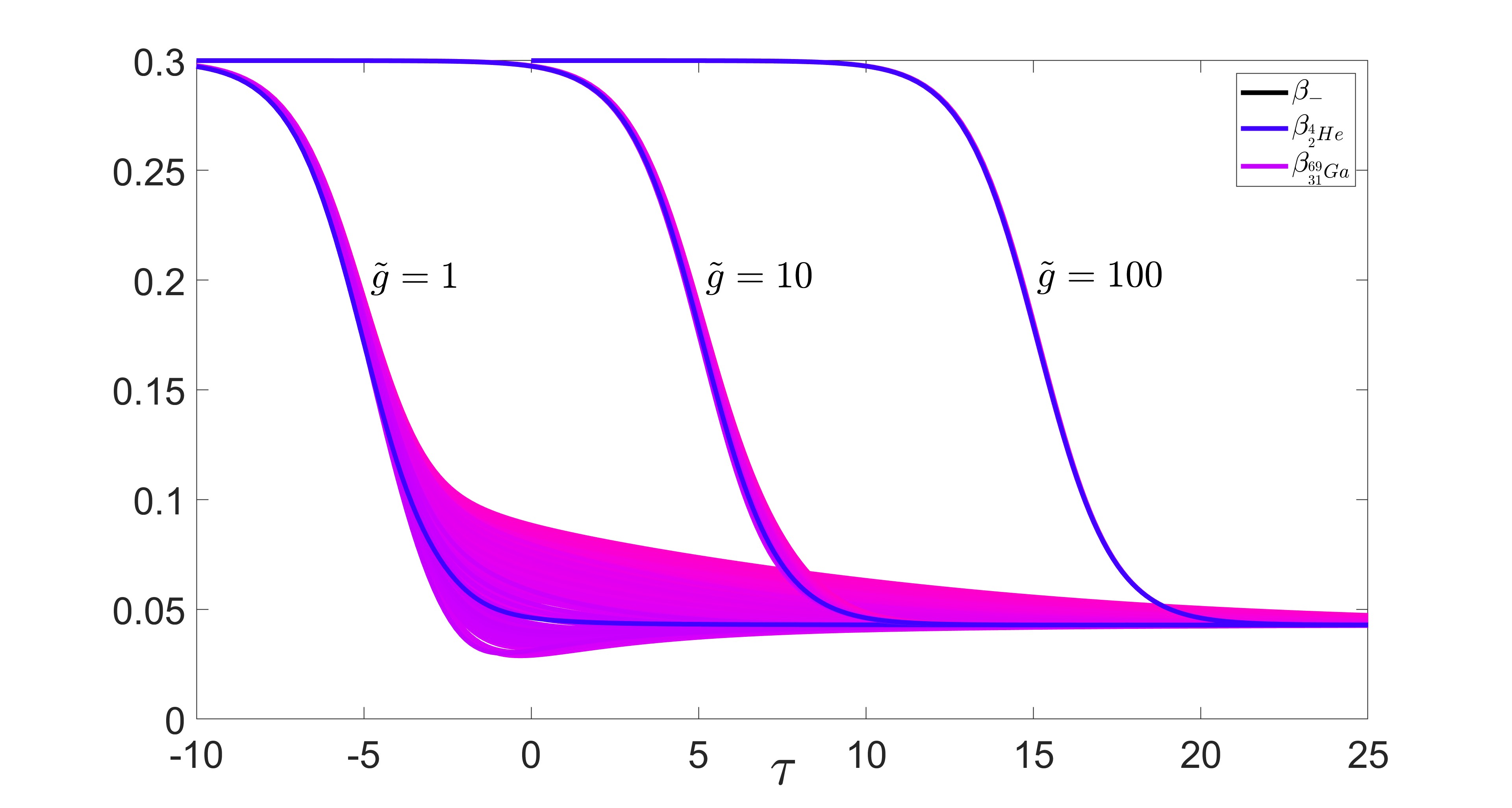}
    \caption{Shock solutions for r-process composition (same as in Fig. \ref{fig:r_process}) with constant friction coefficient of $\Tilde{g}=10^0,10^1$, and $10^2$, as indicated. The different plots are shifted by $\Delta\tau=10$ for visualization purposes.}
    \label{fig:r_process_friction}
\end{figure}

Replacing Eq. \eqref{eq:2} with Eq. \eqref{eq:11}, we obtained new solutions for the shock structure. Results for the $H-He$ and r-process compositions are shown in Fig. \ref{fig:H_He_friction} and Fig. \ref{fig:r_process_friction} respectively. We find no effect on the shock structure up to $\text{\~g}\lesssim10^{-2}$ ($10^{-1}$ for the r-process case). At $\text{\~g}\sim10^{-1}$ ($10^{0}$), the friction force succeeds to couple all the components together on length scales of a few $l_T$, yet enabling ion-ion collisions. For $\text{\~g}>1$ ($10^2$) no velocity differences are obtained and the shock acts as a single fluid.\\

\subsection{Coulomb friction}\label{binary collisions}
Friction via binary Coulomb collisions between different ion species and between ions and electrons was previously considered by \cite{weaver1976structure} based on Spitzer's formula. These interactions strongly depend on the relative velocity between the colliding species as $\propto v^{-3}$, which can be dominated by either the relative drift or thermal motions. Since our model does not contain a self-consistent treatment of photon generation inside the shock, we are unable to determine the temperature along the shock transition layer. 
Instead, we invoke a constant temperature, $k_B T\equiv\theta m_e c^2$, which is treated as a free parameter. We obtain solutions  for a relevant range of 
temperatures found in detailed Monte-Carlo calculations of single fluid shock models  
(see \citealt{ito2020newt,LE2020}).    


The detailed derivation of the Coulomb friction coefficients is presented in appendix \ref{coul_fric_eq}. 
Unlike the constant coefficients of the phenomenological friction model adopted in \S \ref{fric model}, the Coulomb friction
coefficients are characterized by a cubic dependence on the relative velocity of the colliding ions (including the thermal spread) and their atomic charge and mass:
\begin{equation}\label{eq:Coulomb_fric_coef}
     \Tilde{g}_{a a'}=\frac{\mu^2}{\sqrt{2\pi}}\frac{A_a+A_{a'}}{A_a A_{a'}}Z_\alpha^2Z_{a'}^2 \frac{\ln{\Lambda_{a a'}}}{\Delta\beta_{a a'}^3}.
\end{equation}
Here the subscripts $a$ and $a'$ denote either ions (with subscript $\alpha$, atomic number $Z_\alpha$ and mass number $A_\alpha$) or electrons (with $A_e=\mu, Z_e= -1$), 
$\ln{\Lambda_{a a'}}$ is the Coulomb logarithm, given explicitly in Eq. \eqref{eq:app_col-log}, and $\Delta\beta_{a a'}^2=\Delta\beta_{drift,a a'}^2+\Delta\beta_{thermal,a a'}^2$, is the sum of the relative drift, 
$\Delta\beta_{drift,a a'}^2 = (\beta_a -\beta_{a'})^2$, and the thermal speeds, 
$\Delta\beta_{thermal,a a'}^2= 2kT_a/m_a c^2+2kT_{a'}/m_{a'}c^2$.  For simplicity, we shall henceforth assume that all species have
the same temperature, measured in $m_ec^2$ units, $\theta = kT/m_ec^2$, in which case 
$\Delta\beta_{thermal,a a'}^2=2\mu \theta (A_a + A_{a'})/A_a A_{a'}$.  Eq. \eqref{eq:Coulomb_fric_coef} indicates a strong 
dependence on the atomic numbers, implying that the friction force is expected to be much stronger for heavy elements than for light elements.
Shock solutions with Coulomb friction are obtained upon substituting the Coulomb friction coefficients, Eq. \eqref{eq:Coulomb_fric_coef}, into Eq. \eqref{eq:11}.
%

A general expression  for the ratio between the Coulomb friction coupling length, $l_{a a'}$, and the shock length, $l_{sh}= (\beta_u n_e\sigma_T)^{-1}$, is given
in Eqs. \eqref{eq:coul_friction_length} and \eqref{eq:e_coul_fric_length}.
First, we consider Coulomb coupling between ions and electrons in a kilonova. Taking $n_e=10^{25}\,cm^{-3}$ and $\theta=0.1$ as characteristic values, we obtain that for coupling of r-process elements to electrons $\ln{\Lambda_{\alpha e}}\approx 10$, and:
\begin{equation}\label{eq:alpha_e_KN}
\begin{split}
l_{\alpha e,KN} \approx 0.1 \cdot\beta_{u,0.3}^2 \theta_{0.1}^{3/2} Z_{\alpha,50}^{-1} l_{sh},
\end{split}
\end{equation}
The coupling length between electrons and lighter elements, such as He, is longer due to their lower charge:  
\begin{equation}\label{eq:He_e_KN}
\begin{split}
l_{He, e,KN} \approx 3 \cdot\beta_{u,0.3}^2 \theta_{0.1}^{3/2} l_{sh},
\end{split}
\end{equation}
The assumption that the thermal velocity is larger than the drift velocity, which was used to derive Eqs. \eqref{eq:coul_friction_length} and \eqref{eq:e_coul_fric_length}, is always satisfied for coupling with electrons at relevant KN conditions since $\Delta \beta_{thermal,\alpha e} \approx \sqrt{2\theta}\approx0.4\cdot\theta^{1/2}\gtrsim\beta_u$ .

Next we consider the inter-ion coupling length. For a typical r-process nuclei $\alpha$ and $\alpha'$  
we find $\ln{\Lambda_{\alpha\alpha'}}\approx1$ and as long as the relative velocities are dominated by thermal motion, the coupling distance of a heavy nucleus $\alpha$ to the rest of the r-process elements is 
\begin{equation}\label{eq:alpha_alpha_KN}
l_{\alpha,KN} \approx \left(\sum_{\alpha' \neq \alpha} l_{\alpha \alpha',KN}^{-1}\right)^{-1} \approx 10^{-4} \cdot\beta_{u,0.3}^2\theta_{0.1}^{3/2} A_{\alpha,100}^{-1/2} Z_{\alpha,50}^{-1} l_{sh},
\end{equation}
This coupling length holds as long as $\Delta \beta_{drift,\alpha,\alpha'}<\Delta \beta_{thermal,\alpha,\alpha'} \approx 0.001(\theta_{0.1}/A_{100})^{1/2}$ (see discussion below Eq. \ref{eq:coul_friction_length}). If at some point within the shock transition layer the relative velocity, $\Delta \beta_{drift,\alpha,\alpha'}$ exceeds this value the coupling length depends on the drift instead of the thermal velocity, so $l_{\alpha,KN} \propto \Delta \beta_{drift,\alpha,\alpha'}^3$. In such case there is a run-away of the drift velocity and the  coupling breaks. Approximating the drift velocity that develops within the shock transition layer, as long as the thermal velocity dominates over the drift velocity, as $\Delta \beta_{drift,\alpha,\alpha'} \sim \frac{\beta_u l_{\alpha,KN}}{{\rm max}\{l_{sh},l_{\alpha' e,KN}\}}$ we obtain the following criterion for $r-process$ elements with themselves: 
\begin{equation}\label{eq:r_process_couple_crit}
\begin{split} 
\frac{\Delta \beta_{drift,\alpha,\alpha'}}{\Delta \beta_{thermal,\alpha,\alpha'}} &\sim \beta_u \frac{l_{\alpha,KN}}{{\rm max}\{l_{sh},l_{\alpha' e,KN}\}} \left(\frac{A_{\alpha}}{2\mu \theta}\right)^\frac{1}{2}\\&\approx \left\{ 
\begin{array}{ll}
    0.03 \cdot \beta_{u,0.3}^3\theta_{0.1}Z_{\alpha,50}^{-1} & l_{\alpha' e,KN}<l_{sh}  \\
    0.3 \beta_{u,0.3} \theta_{0.1}^{-1/2} & l_{\alpha' e,KN}>l_{sh}
\end{array}
\right. ~,
\end{split}
\end{equation}
where the smallest of the two options always apply.
Following the same argument for coupling of He to heavy ions we obtain $\ln{\Lambda_{He,\alpha'}}\approx10$, and:
\begin{equation}
l_{He,\alpha,KN} \approx \left(\sum_{\alpha' \neq He} l_{He, \alpha',KN}^{-1}\right)^{-1} \approx 0.01 \cdot\beta_{u,0.3}^2\theta_{0.1}^{3/2} l_{sh},
\end{equation}
so the criterion for frictional coupling is 
\begin{equation}\label{eq:He_couple_crit}
\begin{split}
\frac{\Delta \beta_{drift,He,\alpha}}{\Delta \beta_{thermal,He,\alpha}}&\sim \beta_u \frac{l_{He,\alpha,KN}}{{\rm max}\{l_{sh},l_{\alpha' e,KN}\}} \left(\frac{2}{\mu \theta}\right)^\frac{1}{2}\\&\approx \left\{ 
\begin{array}{ll}
    0.75 \cdot \beta_{u,0.3}^3\theta_{0.1} & l_{\alpha' e,KN}<l_{sh}  \\
    0.3 \cdot \beta_{u,0.3} \theta_{0.1}^{-1/2} & l_{\alpha' e,KN}>l_{sh}
\end{array}
\right. ~.
\end{split}
\end{equation}

\begin{figure}
    \centering
    \includegraphics[width=\columnwidth]{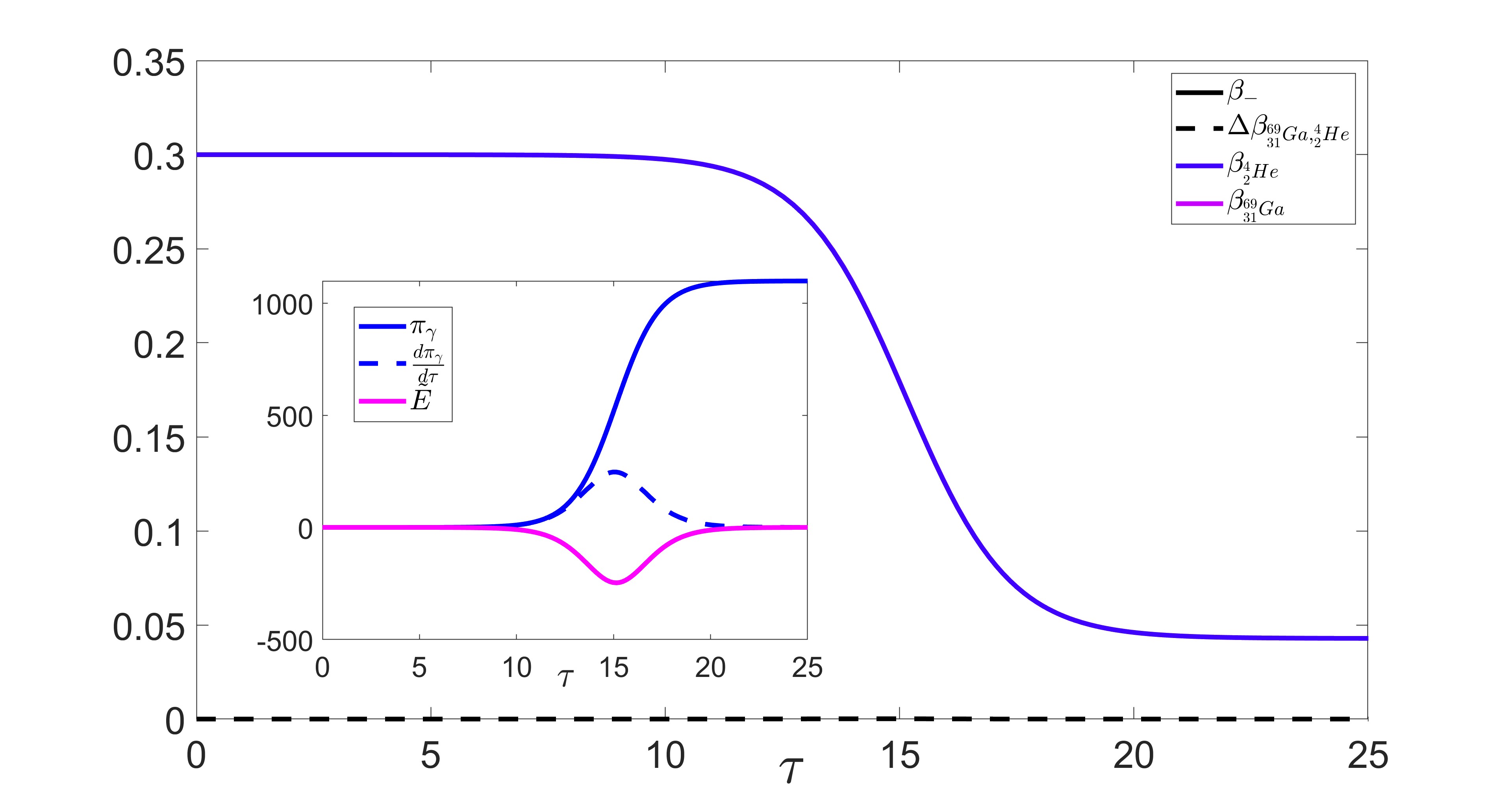}
    \caption{Shock structure for r-process composition with $\beta_u=0.3$, $n_e=10^{25}$ and $\theta=0.2$. 
    The strong coupling of all ions by Coulomb friction is clearly seen main panel.}
    \label{fig:rp_coul_fric}
\end{figure}

Eqs.  \ref{eq:alpha_e_KN}, \ref{eq:He_e_KN}, \ref{eq:r_process_couple_crit} and \ref{eq:He_couple_crit} provide the coupling criteria of the various plasma constituents when only friction is at work. In reality, electrostatic field, as well as anomalous coupling, may also play a role. Applying these criteria to shocks in KN ejecta we find that the heavy r-process elements are tightly coupled and therefore behave as a single fluid (Eq. \ref{eq:r_process_couple_crit}). The heavy element fluid is coupled to electrons via Coulomb friction (Eq. \ref{eq:alpha_e_KN}) with the assistance of electrostatic field. Finlay, while the Coulomb coupling of He to the rest of the plasma elements is marginal, the electrostatic field guaranties strong coupling of the He as well. The result is that for the conditions expected in KN ejecta we expect all the plasma constituents to behave as a single fluid, at least as long as no pairs are present (see \S\ref{pairs_role} for the effect of pairs).
This expectation is verified by detailed calculations (see Fig. \ref{fig:rp_coul_fric} for example).

\begin{figure}
    \subfloat[Shock structure for multi-ion solar-like ambient, including coulomb interactions, with $\beta_u=0.4$, $n_e=10^{15}cm^{-3}$ and $\theta=0.2$. The left panel shows the electric field and radiation pressure evolution. The right panel depicts the velocity profile of the different constituents. A relative drift of $\sim0.15c$ is obtained giving rise to collision energies of $\sim35-40MeV$.]{ \centering   \includegraphics[width=\columnwidth]{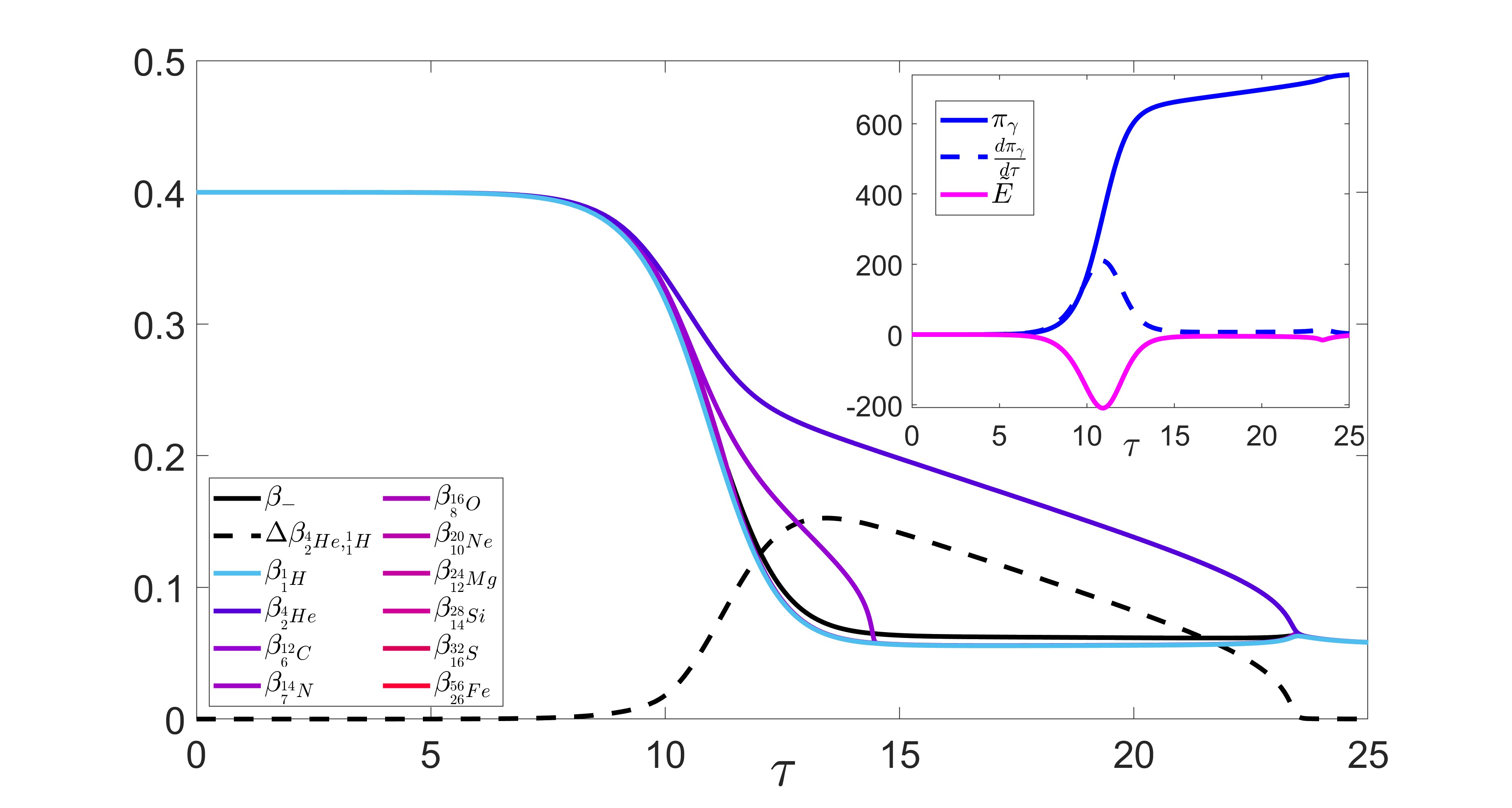}}
    \hfill
    \subfloat[Non-elastic collision depth for different ions with $_2^4He$ within the shock. Cross sections are taken to be geometrical, and the length is calculated above the fusion barrier as explained below (c). roughly $\sim0.6$ of all ions would interact with $He$ ions.]{
    \centering    \includegraphics[width=\columnwidth]{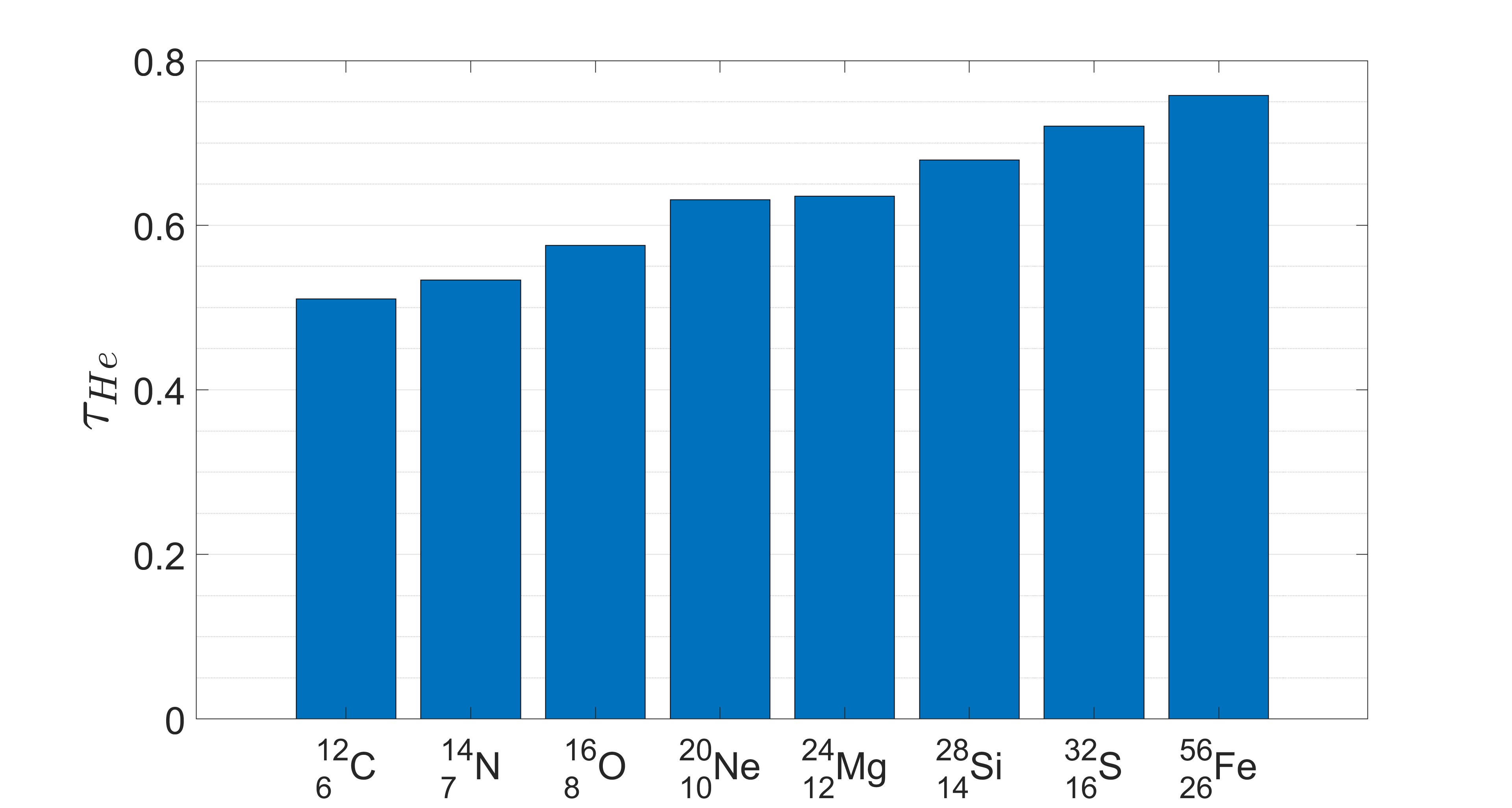}}
    \hfill
    \subfloat[Non-elastic cross-sections for $Fe$, $Si$ and $N$ with $He$ taken from \url{https://www-nds.iaea.org/exfor/endf.htm}. The geometrical cross-sections used in our estimations are calculated by $\sigma_{X(He,non)}=0.0452\cdot(A_X^{1/3}+A_{He}^{1/3})^2\,barn$ and shown by the horizontal dashed lines in the matching colors. The fusion barriers we use are shown in the matching vertical dotted lines.]{
    \centering    \includegraphics[width=\columnwidth]{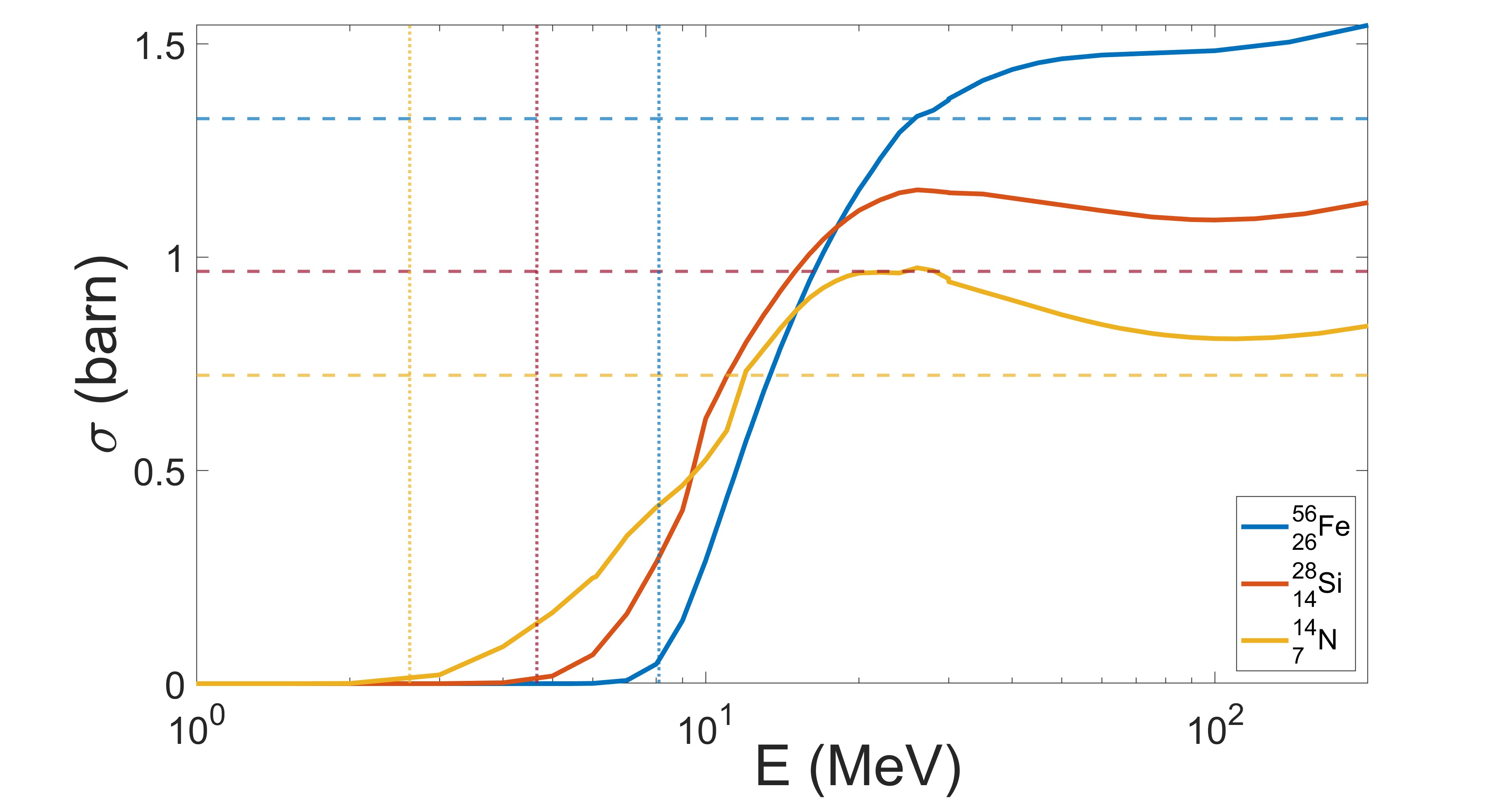}}
    \hfill
    \caption{}
\label{fig:Solar_th02_tau}
\end{figure}

For a supernova, taking $Z=\{1,2\}$, $n_e=10^{15}\,cm^{-3}$ and $\theta=0.01$, we find $\ln{\Lambda_{\alpha\alpha'}}\approx\ln{\Lambda_{\alpha e}}\approx 10$, and
\begin{equation}\label{eq:SN_l_fric}
\begin{split}
&l_{\alpha\alpha',SNe}\approx10^{-3}\beta_{u,0.1}^2\theta_{0.01}^{3/2} l_{sh} , \\
&l_{\alpha e,SNe}\approx0.01\cdot\beta_{u,0.1}^2\theta_{0.01}^{3/2} l_{sh}.
\end{split}
\end{equation}
As seen, ion-ion collisions dominate the Coulomb friction force in this case as well and the electrostatic field guaranties coupling to the electrons when the ions behave as a single fluid. Equation \eqref{eq:SN_l_fric} indicates that strong coupling is 
anticipated at not-too-high velocities.  In fast enough shocks, $\beta\gtrsim 0.4$, where the temperature inside the shock exceeds 
$\sim 100$ keV \citep{ito2020newt,Levinson_2020}, Coulomb friction alone may be insufficient to prevent velocity spread behind the shock (as we show later the formation of pairs at these temperature is expected to further reduce the coupling).
An example of a shock solution with  $\beta_u = 0.4$ and $\theta_e=0.2$ is shown in Fig. \ref{fig:Solar_th02_tau}. 
A relatively large drift between $He$ and all other ions is observed (e.g. $Fe$, $Si$). The generation of plasma turbulence by ion beam instability, expected under such conditions, can accelerate particles and alter the breakout signal.  
The anomalous friction due to plasma turbulence will probably reduce the drift velocity, yet it may still be sufficient to ignite helium-induced $\gamma$ emission as the shock passes through the star ($He+X\longrightarrow He'+X'+\gamma$), and conceivably (though with a lower cross-section)  alpha-capture ($He+ _Z^A X\longrightarrow _{Z+2}^{A+4} X+\gamma$). 


\subsection{Thermal effects}
There are two temperature-related aspects, which we do not address in this work, that are predicted to contribute to the decoupling of the ions, and thus to the collision rate. First, since $l_{\alpha,\alpha'}<l_{\alpha,e}$ by a factor of $\mu$, the ion temperature can potentially be higher than that of the electrons and the radiation along the shock transition layer. The reason is that the heating of the ions due to Coulomb collisions that are responsible for keeping them coupled within the shock transition layer (although each ion species decelerates at a different rate due to the electrostatic field), is done over $l_{\alpha,\alpha'}$. The transfer of this heat to the electrons and then to the photons, which dominates the heat capacity of the shock transition layer and the downstream, is done over $l_{\alpha,e}$. Thus, the ions and the electrons can have two distinct temperatures, where the former can, in some scenarios, be significantly higher than the latter. Increasing temperature will decrease the efficiency of inter-ion binary collision, increasing the coupling length.
A second thermal effect is due to the thermal velocity dispersion within each beam.  The collision rate is a function of the coupling length $l\propto E^{3/2}$ and the density, which, assuming a  Maxwellian distribution for each ion species, is proportional to $\propto \sqrt{E}e^{-E/E_0}$. Integrating altogether, this effect gives a factor of $\sim2$ to the total rate. We leave this, again, to future work.

\section{The role of pairs}\label{pairs_role}

For shock temperatures exceeding $\sim50$ keV ($\theta\sim0.1$) pair production starts to play a role. This will be important at shock velocities  $\beta_u \gtrsim0.35$ for $n_e\sim10^{15}$ cm$^{-3}$ and light element composition, and $\beta_u \gtrsim0.25$ for $n_e\sim10^{25}$ cm$^{-3}$ and heavy r-process element ejecta, corresponding to SNe and kilonova environments respectively (\citealt{LE2020}). 
The presence of positrons with larger than unity multiplicity can substantially affect the shock structure.  First, positrons tend to decelerate 
faster than all other species since, in contrast to electrons, the electric and radiation forces acting on them are in the same direction, and they have a large charge to mass ratio, $\sim2000$. 
Thus, in the absence of tight coupling by friction, the positrons are strongly compressed and decouple from the other species early on, as
shown by detailed calculations for relativistic RMS \citep{Levinson_2020}. Consequently, plasma instabilities should play important role and need to be taken into consideration.   
When the coupling between electrons and positrons by Coulomb friction is strong enough, 
complete screening of the electric field inside the shock ensues, considerably affecting the dynamics of alpha particles (see details below).  
Second, the larger the pair multiplicity, the larger the opacity for Thomson scattering, resulting in a smaller physical shock width. This can potentially alter the 
ratio between the radiation length and the coupling length of various ions. 

In order to accurately quantify these effects, self-consistent calculations of pair loaded multi-ion RMS is needed.  This, in turn, requires a proper account of photon production processes
in the model, which is beyond the scope of our current analysis.  We leave the construction of a generalized model for future work. 
In what follows, we study the effect of pair loading using a simplified model.  To be concrete, we assume that the upstream plasma contains electron positron pairs with 
a fixed multiplicity, $M\equiv n_{+,u}/(n_{e,u} - n_{+,u})$, where $n_{+,u}$ is the upstream positron density and $n_{e,u} = n_{+,u} + \sum_\alpha n_{\alpha,u} Z_\alpha$ is the total electron density, and ignore pair production inside the shock.  With this definition we have $j_+/j_e = M/(M+1)$, where $j_+= n_+\beta_+$ (see next subsection).
We explore solutions for a multiplicity range $M = 1- 30$, as indicated by Monte-Carlo simulations for fast Newtonian RMS \citep{ito2020newt}, taking into
account Coulomb friction between all species.  The corresponding equations are derived in appendix \ref{app_sec_pairs}.

%

\subsection{A simplified pair loaded multi-ion RMS model}\label{pairs_model}
Since our model ignores pair creation, the positron flux is conserved, $j_+ = n_+ \beta_+ =$ const.  The charge neutrality condition 
then generalizes to 
\begin{equation}\label{eq:pairs_flux_cons}
    \sum_{\alpha}j_{\alpha}Z_{\alpha}+j_{+}=j_{e}.
\end{equation}
Since the radiation is coupled to positrons in addition to electrons, the advection of radiation in the shock depends on both $\beta_e$ and $\beta_+$, and
Eq. \eqref{eq:app_Tox} is no longer valid.  Henceforth, we assume that the advection velocity of the radiation field is given by 
$\beta_{adv} = (n_e\beta_e + n_+\beta_+)/(n_e+n_+)$.  As will be shown below, at multiplicities $M \gtrsim$ a few, the electron and positron fluids 
are strongly coupled, so that practically $\beta_{adv} = \beta_e$.  The reason is that for $M>1$ the electric field inside the shock is nearly completely 
screened, hence both fluids experience the same force (the radiation force) and, consequently, the same deceleration.  With this choice of $\beta_{adv}$
the energy flux of the radiation field is given by Eq. \eqref{eq:radiation_E_flux_pos}, where $\tau$ is the optical depth contributed by the electrons, that is,
$d\tau = \sigma_T n_e dx$.

Now, in the momentum equations, one should take into account the fact that the momentum transferred to the electron fluid by the radiation at any given position 
is proportional to the relative number of scatterings, $n_e/(n_e+n_+)$, and likewise for the positron fluid. Moreover, the friction forces associated 
with the positron fluid should also be accounted for.  The resultant momentum equations are given in Eqs. \eqref{eq:app_momentum_ep}-\eqref{eq:app_momentum_ion}. 
For the results presented below we assume that friction forces are solely due to binary Coulomb collisions, as in Sec. \ref{binary collisions}, but including
the positron fluid. Specifically we add ion-pairs bilateral friction and present a new electron-positron friction given by Eq. \eqref{eq:Coulomb_fric_coef} 
with $A_\pm=\mu$ and $Z_\pm=\pm1$.

\begin{figure}
    \subfloat[Shock structure for pair-loaded r-process plasma with $M=10$, $\theta=0.2$ and $\beta_u=0.3$. The helium mass fraction is 0.04, meaning one alpha particle per each heavy ion. For comparison with previous cases, the various profiles are plotted as functions of the pair unloaded optical depth, $d\tau^\star = \frac{n_e}{1+M}\sigma_T dx.$]{ \centering   \includegraphics[width=\columnwidth]{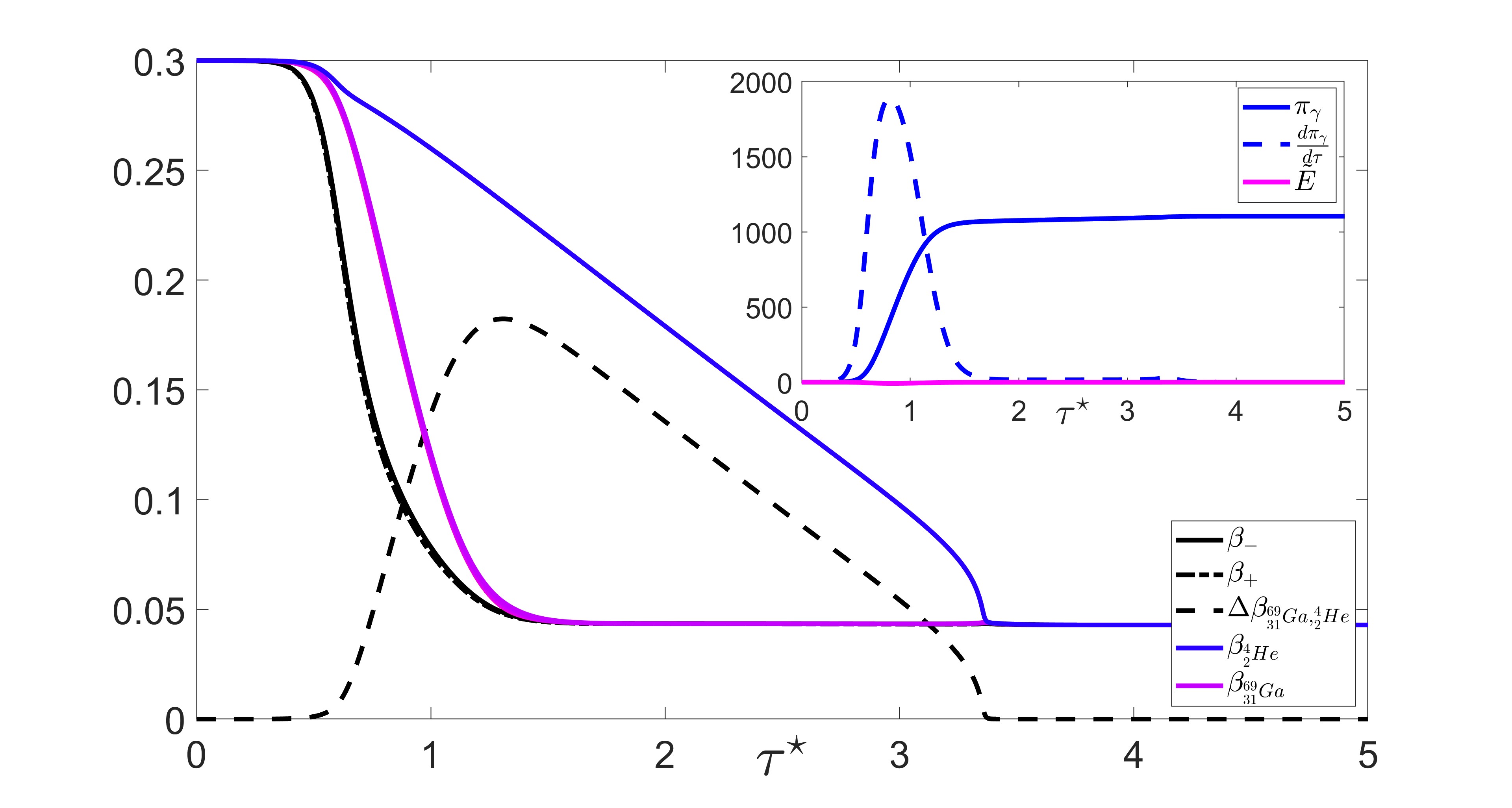}}
    \hfill
    \subfloat[Non-elastic collision depth for different heavy ions with $_2^4He$ within the shock. Cross sections are taken to be geometrical, and the length is calculated above the fusion barrier. Roughly $\sim0.1$ of all heavy ions would interact with $He$ ions. Increasing the helium mass fraction to $\sim0.1$ will increase the $\tau_{He}$ respectively to $\sim0.25$.]{
    \centering    \includegraphics[width=\columnwidth]{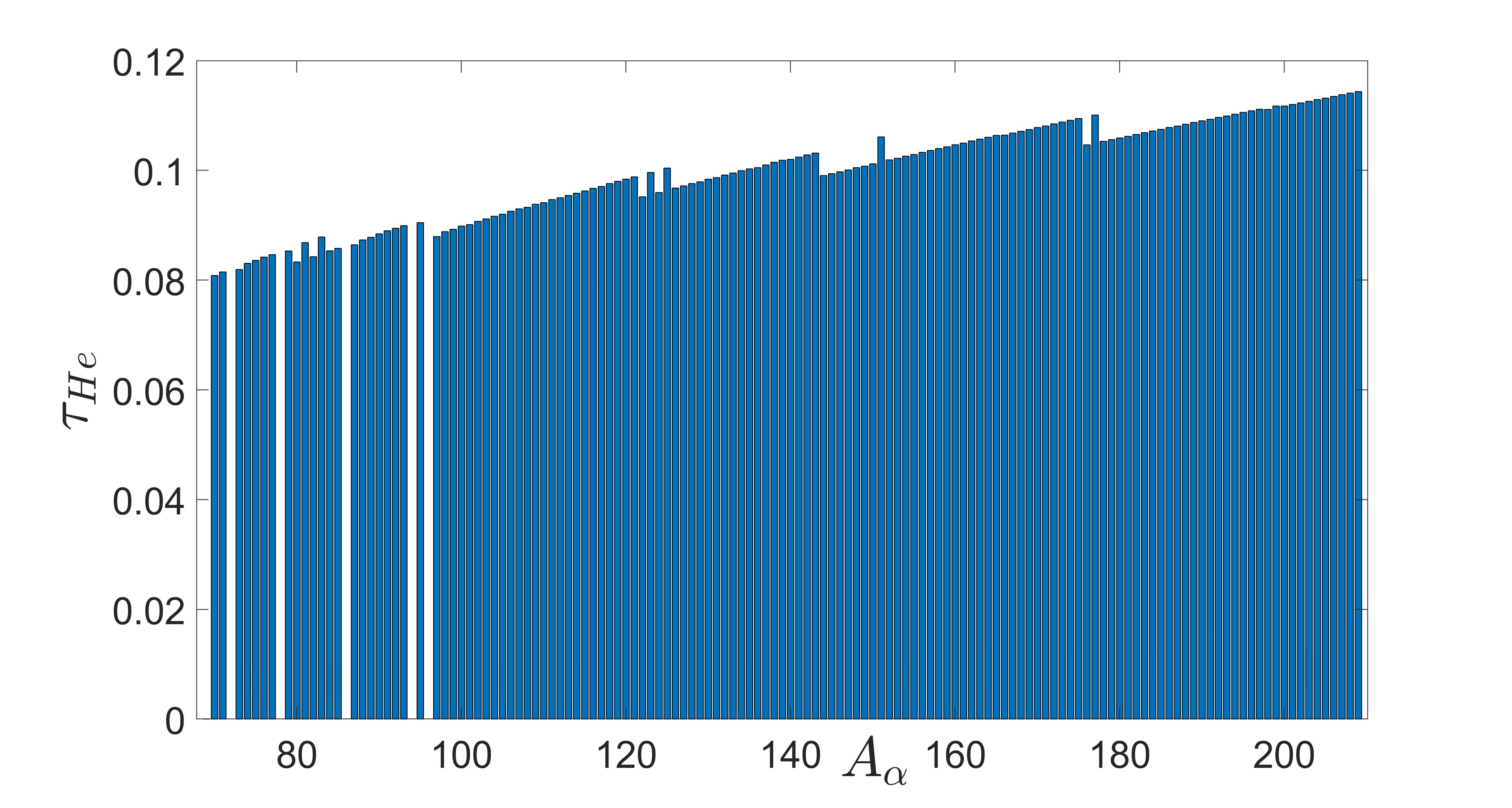}}
    \hfill
    \caption{}
\label{fig:R_process_M10}
\end{figure}

\begin{figure}
    \centering
    \includegraphics[width=\columnwidth]{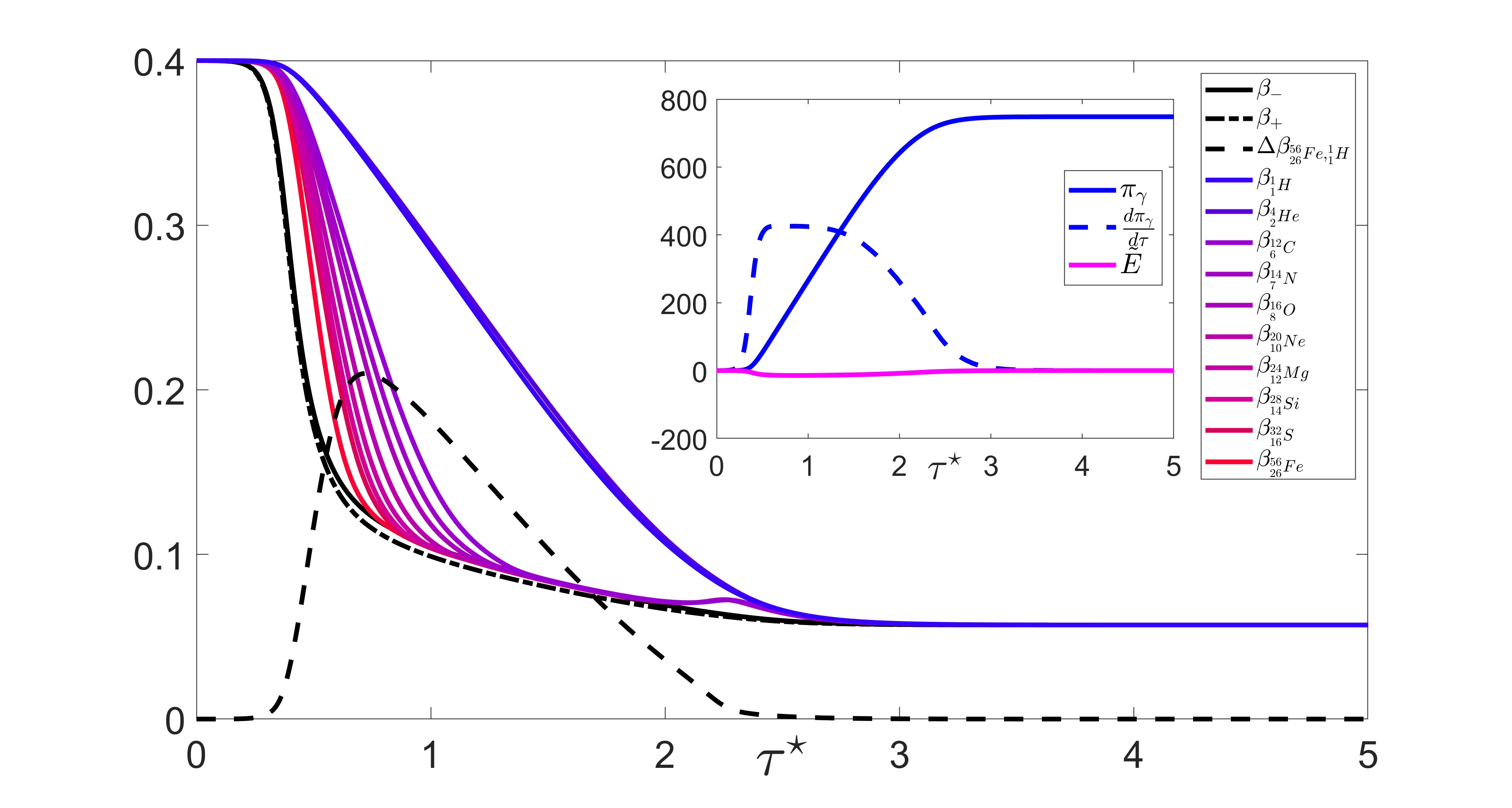}
    \caption{Shock structure for pair-loaded solar-like plasma with $M=10$, $\theta=0.2$ and $\beta_u=0.4$....}
    \label{fig:Solar_M10}
\end{figure}

Two examples of pair loaded shock solutions are exhibited in Figs. \ref{fig:R_process_M10} and \ref{fig:Solar_M10}.  For comparison purposes the 
velocity profiles in these figures are plotted as functions of the pair unloaded optical depth, $d\tau^\star = \frac{n_e}{1+M}\sigma_T dx$.  Fig. \ref{fig:R_process_M10}a 
shows a solution of a shock with $M=10$ propagating in r-process ejecta under the same conditions as in Fig. \ref{fig:rp_coul_fric}. As seen, while in the absence 
of pairs all species, including He ions, are fully coupled (Fig. \ref{fig:rp_coul_fric}), in the presence of pairs the He ions develop a large relative drift with respect to the remaining ion cluster (represented by the magenta line in Fig. \ref{fig:R_process_M10}a). This result can be understood by considering the effect of pairs on the ratio between the Coulomb friction length scale and the shock width. 
The shock width shrinks as $\l_{sh} \propto M^{-1}$, 
and the Coulomb length scales for coupling of ions and electrons (as well as positrons) are also reduced by the same factor ( $ M^{-1}$). Consequently,  Eqs. \eqref{eq:alpha_e_KN} and \eqref{eq:He_e_KN} are unaffected by the presence of pairs. Likewise, the ion-ion Coulomb length scale is not affected directly by pairs, however, the reduction of the shock width implies that the r.h.s in Eqs. \eqref{eq:alpha_alpha_KN}-\eqref{eq:He_couple_crit} is multiplied by $M$. Similarly, in Eq. \eqref{eq:SN_l_fric} the expression for $l_{\alpha\alpha',SNe}$ is multiplied by $M$ on its r.h.s, while the expression for 
$l_{\alpha e,SNe}$ is unaffected by pairs.

We can now understand why the presence of pairs leads to the decoupling of the He.  First, at the anticipated multiplicity, pairs screen nearly completely the electrostatic field, hence, in the absence of significant anomalous friction, the coupling is solely due to Coulomb friction. Eq. \eqref{eq:alpha_e_KN} implies that for mildly relativistic shocks the heavy ions are at least marginally bound to the pairs and therefore they are expected to follow the electrons either tightly or with a short lag. The He ions, however, are not bound to the electrons (Eq. \ref{eq:He_e_KN}) and for $M \gtrsim 5$ are also not expected to be bound to the heavy ions. We therefore expect that if $M \gtrsim 5$ within the shock transition layer then a significant drift velocity is developed between He and the the heavy ions.


The separation of the He ions in Fig. \ref{fig:R_process_M10}a is large enough to allow inelastic collisions with the heavy ions downstream.  The optical depth 
for collisions above the Coulomb barrier of different ions is shown in the lower panel Fig. \ref{fig:R_process_M10}b for He mass fraction of $X_{He} = 0.04$. We find, as expected,
that the probability for inelastic collisions is proportional to $X_{He}$. For $X_{He} = 0.1$, the highest value found in some simulations \citep{Tarumi2023}, our simplified 
model indicates that about 25\% of the r-process isotopes will be converted through inelastic collisions with He ions in a shock moving at a velocity $\beta_u=0.3$. The 
dependence of the He velocity separation on temperature and pair multiplicity necessitates a self-consistent treatment of pair-loaded shocks in order to assess more 
accurately the conditions under which significant nuclear transmutations by collisions of heavy ions with alpha particles is expected. 
We defer such analysis for future publication.

\section{The role of free neutrons}\label{sec:free_neutrons}
\begin{figure}
    \centering
    \includegraphics[width=\columnwidth]{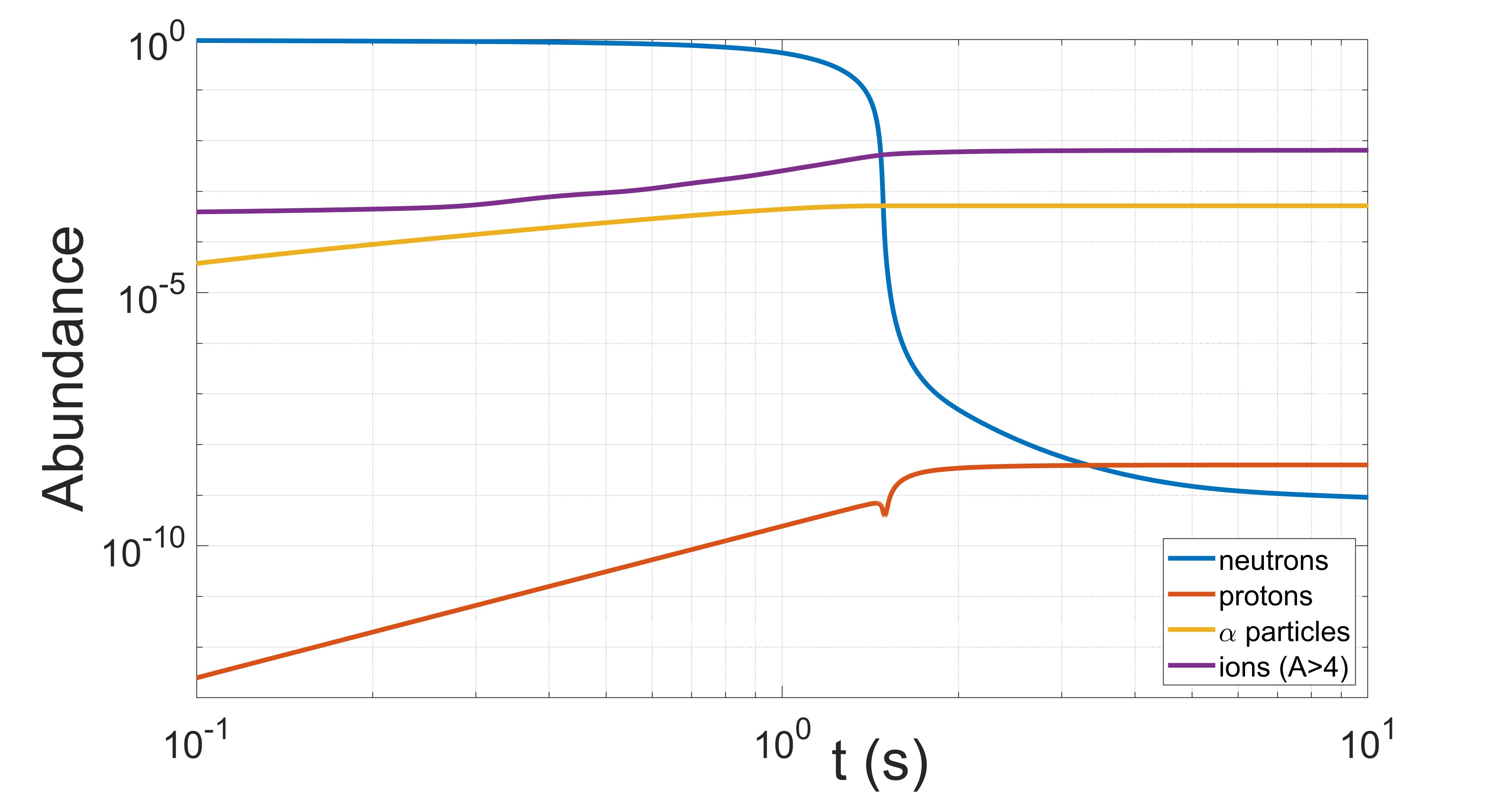}
    \caption{Abundance temporal evolution of the different constituents in a kilonova remnant around the "freeze-out" phase of the r-process (here freeze-out takes place roughly at 1.5s). The results were obtained using the "Sky-Net" r-process module \citep{Lippuner2017}.The initial conditions are $Y_e=0.01$, $T_0=6GK$, $s=10\frac{k_B}{baryon}$ and $\tau_{exp}=7.1ms$.}
    \label{fig:rp_abundance}
\end{figure}

The ejected material in BNS merger consists of various r-process isotopes, alpha particles, and free neutrons.
The rapid neutron capture process lasts for about $1$ sec before the mixture reaches freeze-out, followed
by the decay of unstable isotopes into their final stable states, whereby the final composition of the r-process ejecta is gradually established. 
An example of the evolution of the r-process mixture is shown in Fig. \ref{fig:rp_abundance}.  

If the shock crosses the ejecta early enough, a relatively large abundance of free neutrons may be present in the upstream region.  
One might naively suspect that free neutrons advected with the upstream flow will cross the shock undisturbed, and might trigger fission via various types of inelatic collisions with the shocked ions.
However, there exist three main channels through which neutrons crossing the shock 
may interact with the ions inside the shock: elastic scattering $(nX,n'X')$, inelastic radiative scattering $(nX,n'X'\gamma)$ and other non-elastic scattering processes, most of which results in nuclear transmutation (e.g., neutron capture, spalation, fission, etc.). A reaction that can induce nuclear transmutations requires the collision energy to exceed the activation barrier ($\lesssim 25$ MeV for most isotopes), while
elastic collisions are effective also at low energies. Thus, if the elastic collision rate is shorter than the shock crossing time, it can couple the neutrons to the ions,
preventing large velocity spreads.

\begin{figure}
    \centering
    \includegraphics[width=\columnwidth]{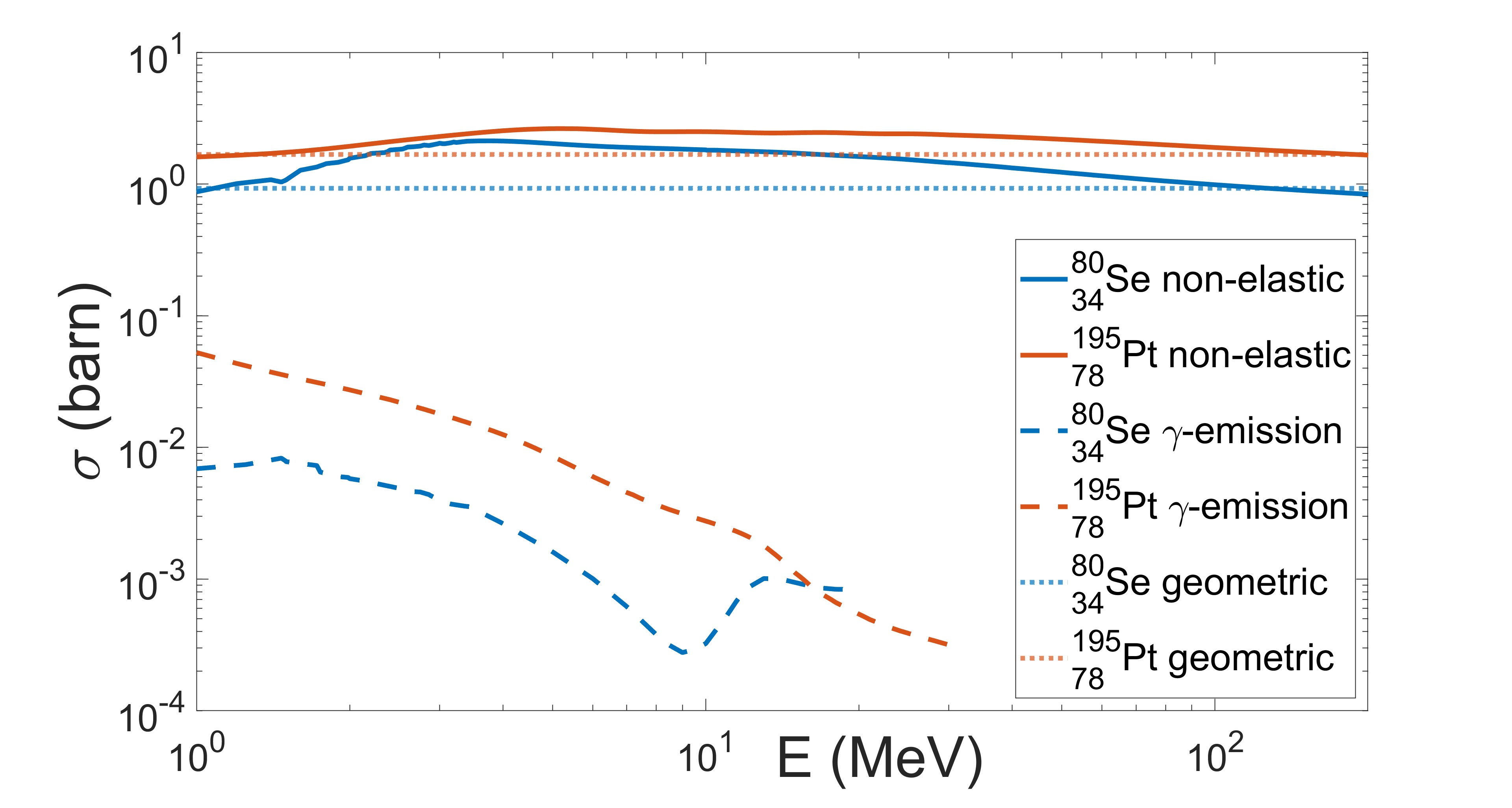}
    \caption{Total non-elastic, $\gamma$-emission and calculated geometric cross-sections for two representative isotopes ($_{34}^{80}Se$ and $_{78}^{195}Pt$) with incoming neutron for $1-200MeV$. $\gamma$-emission is only $\sim1\%$ of total inelastic interactions, meaning ionic transmutations are probable. Data is taken from Evaluated Nuclear Data File (ENDF): \url{https://www-nds.iaea.org/exfor/endf.htm}.}
    \label{fig:non_el_cs}
\end{figure}

Figure \ref{fig:non_el_cs} shows the geometric (an approximation of the eleastic), non-elestic and $\gamma$-emission cross-section. It shows that the latter is smaller by about two orders of magnitude compared to the other two, which are of the same order. 
In what follows, we explore whether elastic collisions can couple the neutrons to the ions within the shock transition layer, thereby avoiding large drift velocity between them, or whether the neutrons cross the shock almost unaffected. 
%

\subsection{Elastic collisions, friction force and the effect on the shock structure}
The mean momentum exchanged along the flow direction in an elastic collision between
a neutron of mass $A_nm_p$, with $A_n=1.0013$, and an ion of mass $A_\alpha m_p$ is:
\begin{equation}
    <\Delta p_{n\alpha}> = -\frac{A_\alpha A_n}{A_\alpha +A_n} m_p \Delta\beta_{n\alpha}\cdot c,
\end{equation}
 Here we assumed for simplicity a point-like neutron scattered off a spherical ion with a uniform differential cross section ($\frac{d\sigma}{d\Omega}=const$). Taking $r_{nucleon}\sim 1.2\,fm$, the total cross-section is approximated by $\sigma_{n\alpha} = \pi r_{nucleon}^2 = \pi\cdot 1.2^2\,fm^2\cdot A_\alpha^{2/3} = 4.5\cdot 10^{-2} A_\alpha^{2/3}\sigma_T$.

The resultant force per unit volume acting on the neutron fluid by ion beam of type $\alpha$ is:
\begin{equation}
\mathcal{F}_{n\alpha}= n_{n} \nu_{n\alpha}<\Delta p_{n\alpha}>,
\end{equation}
where $\nu_{n \alpha}$ is the elastic scattering rate given by:
\begin{equation}
\nu_{n\alpha}=n_{\alpha}\sigma_{n \alpha}|\Delta\beta_{n\alpha}\cdot c|.
\end{equation}
Momentum conservation readily implies $\mathcal{F}_{\alpha n}= - \mathcal{F}_{n\alpha}$.
Since $\sum_\alpha\mathcal{F}_{n\alpha}=A_{n} m_p n_{n} \beta_{n} n_e \sigma_T \frac{d\beta_{n}}{d\tau} c^2$, one finds:

\begin{equation}\label{eq:el_n_ion_eom}
\begin{split}
    \frac{d\beta_{n}}{d\tau} &= -4.5\cdot10^{-2}\cdot \sum_\alpha\frac{1}{\beta_n}\frac{n_{\alpha}}{n_e}\frac{A_\alpha^{5/3}}{A_\alpha+A_n}|\Delta\beta_{n\alpha}|\Delta\beta_{n\alpha} \\
    &= -4.5\cdot10^{-2}\cdot \sum_\alpha \frac{\beta_e}{\beta_n}\frac{\Tilde{j}_{\alpha}}{\beta_{\alpha}}
    \frac{A_\alpha^{5/3}}{A_\alpha+A_n}|\Delta\beta_{n \alpha}|\Delta\beta_{n \alpha}.
\end{split}
\end{equation}
%
The force exerted on the ion fluid $\alpha$ by the neutrons can be added to Eq. \eqref{fric model}, with
the friction coefficient given by:
\begin{equation}\label{eq:g_neutrons}
    \Tilde{g}_{\alpha n}\equiv4.5\cdot10^{-2}\frac{A_n A_\alpha^{5/3}}{A_\alpha+A_n}|\Delta\beta_{n,\alpha}|.  
\end{equation}

\begin{figure}
    \centering
    \subfloat[]{\centering\includegraphics[width=\columnwidth]{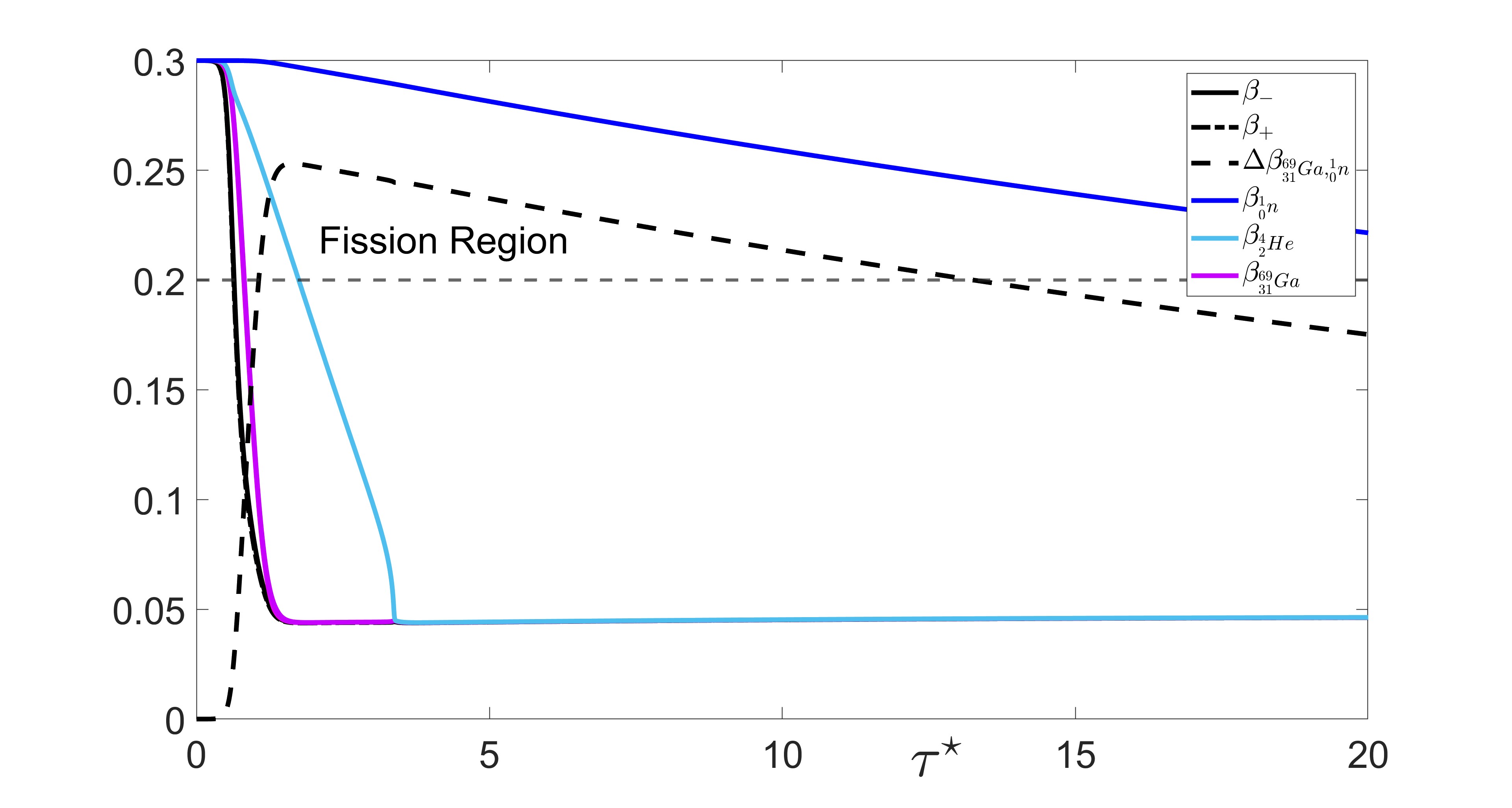}}
    \hfill
    \subfloat[]{\centering\includegraphics[width=\columnwidth]{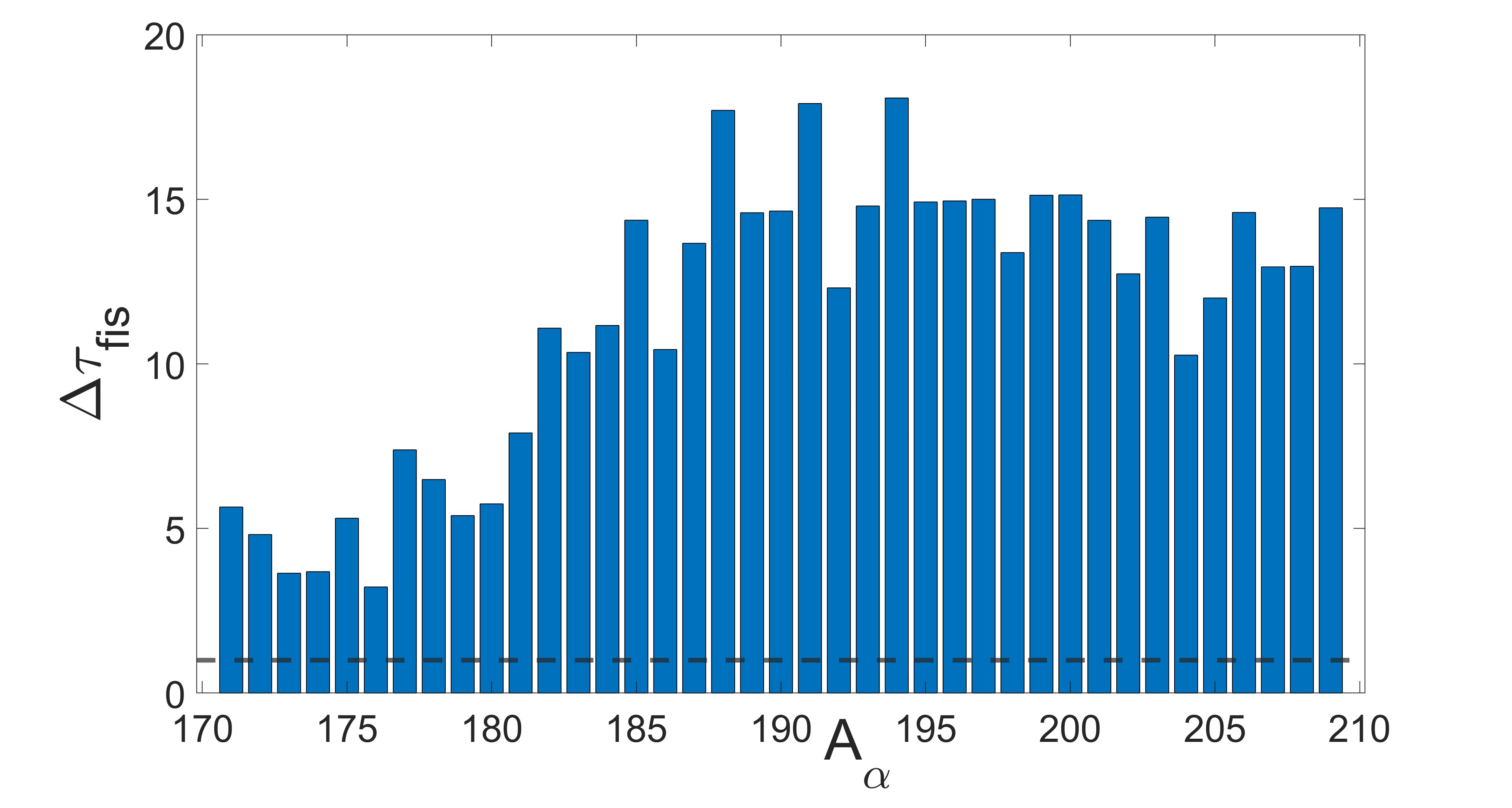}}
    \hfill
    \caption{(a) Shock profile moving through r-process ambient with $69\leq A \leq209$, $n_e=10^{25}cm^{-3}$ and $\Tilde{j}_n=1$ at  $\beta_u=0.3$. Energies higher than the fission barrier are obtained within the mentioned region. All charged particles are coupled, where the $_{31}^{69}Ga$ red line represents their profile. (b) Number of collisions for heavy ions above the fission barrier region. Fission barriers, based on the finite-range liquid drop model (FRLDM), are taken from \citealp{fission_bar}. Cross-sections are taken to be geometrical.}
    \label{fig:neutrons_03_j_1}
\end{figure}

Fig. \ref{fig:neutrons_03_j_1} depicts the shock structure and neutron-heavy ions fission rates, including free neutrons, for the same parameters used in Fig. \ref{fig:R_process_M10}. We added free neutrons with $\Tilde{j}_n=1$ (corresponding to the pre-freeze-out region in the r-process case), which means we have about 10 neutrons per ion, giving significant fission rates. Reducing the neutrons fraction to $\sim1$ neutron per ion ($\Tilde{j}_n=0.1$), for the same shock properties, will linearly reduce the fission rate to the critical value of 1. Note that at earlier times, density and temperature may be higher ($\theta\sim0.2$ and $n_e\gtrsim10^{27}cm^{-3}$). This, however, does not alter the conclusions, showing similar results to fig. \ref{fig:neutrons_03_j_1}.  

\subsection{Neutron-induced transmutation}
We have seen that elastic collisions cannot couple the neutrons to the ions and the electrons. Hence, in the frame of the shock downstream, which has a sub-MeV temperature and a velocity of $(6/7)\beta_u$ compared to the upstream, the neutrons have a kinetic energy of about $0.4\cdot m_p \beta_u^2 \approx 40 \beta_{u,0.3}^2$ MeV. This exceeds the activation energy for most interactions. The effect of the sudden "heating" of the neutrons during times at which the nucleo-synthesis starts to fall out of equilibrium, but before freeze-out, is beyond the scope of this work. Yet we expect it to be significant.  

\section{Application to BNS mergers}\label{sec:BNS_mergers_applc}

The nucleosynthesis of r-process elements in the outflow of BNS mergers takes place during the first second or so following its ejection, as the neutron-rich material decompresses. The exact evolution of the thermodynamic conditions as well as the nuclear reaction chian within each fluid element depends on its initial state and expansion velocity (e.g., \citealt{Korobkin2012}). As long as nuclear reactions (neutron capture, fission, etc.) take place, the temperature of the gas remains roughly 50 keV and the free neutrons are over-abundant (dominate the number density). A sharp transition in the thermodynamic evolution takes place when the reactions terminate, roughly a second after the ejection at a density of about $100 ~{\rm gr~cm^{-3}}$. In the first stage, almost all of the free neutrons are captured over a very short time scale (much shorter than a second). Subsequently, the density and temperature evolve as in a freely expanding radiation-dominated gas, $\rho \propto t^{-3}$ and $T \propto t^{-1}$. The composition during this stage evolves only by radioactive decay towards the valley of stability. As we discuss below, this evolution implies that the properties of a shock that is driven into the ejecta strongly depend on whether it propagates in the ejecta before or after nuclear reactions freeze out. 

Several different shocks are expected to cross the sub-relativistic material ejected from a BNS merger. First, in the commonly accepted scenario, and as supported by observations of GW170817, a jet is launched from the compact remnant formed in the aftermath of BNS coalescence and propagates through the merger ejecta. The motion of the jet inflates an expanding cocoon that drives an asymmetric RMS in the ejecta.  The shock velocity is fastest along the jet axis (can approach the speed of light) and decreases with increasing angle. 
The angular distribution of the shock velocity depends on the density distribution of the ejecta, as well as the jet luminosity and opening angle, and is not known exactly. However, simulations suggest that it can exceed $0.1 c$ and even approach $0.3 c$ over a significant range of angles \citep{Gottlieb2022}. Overall, a small fraction (of an order of a few percent) of the ejecta is shocked by the jet and the cocoon. This material is expected to dominate the early KN emission, on time scales of minutes to hours \citep{Gottlieb2018a,Hamidani2023,hamidani2022}. The shock can cross the ejecta both before or after the nucleosynthesis freeze-out and affect the composition of the cocoon material. A significant change in the composition of this material might affect the properties of the early kilonova signal.  

A second source of shocks is the interaction between various components of the ejecta itself. There are several different mechanisms that expel matter, each operates on a different time scale after the merger and produces outflow with a different range of velocities (see, e.g., \citealt{nakar2020} and references within). Since some of these produce relatively fast ejecta over long time scales, shocks are expected to form when different components of the outflow collide. For example, \cite{Nedora2021} find a fast (0.15-0.3 c) spiral-wave driven wind that lasts as long as the central remnant does not collapse to a black hole. This fast material, which may be ejected for a second or longer, is expected to interact with slower material, driving shocks at a velocity of $\sim 0.1-0.2$ c. Observations of GW170817 reveal a comparable amount of mass in ejecta having a velocity of 0.2-0.3 c and ejecta with a velocity of $0.1-0.2$ c. It is also likely that an even faster material was ejected. It is plausible that collisions during the ejection of this fast component generated shocks with velocities in the range $\sim 0.1-0.2$ c, and possibly even faster. These types of shocks form during the main mass ejection episode and, therefore, most likely cross the ejecta before the nucleosynthesis freeze-out, although it is conceivable that some shocks form also at later times.

Finally, the central remnant may also drive a quasi-spherical, or a wide-angle, shock into the ejecta. For example, a long-lasting magnetar can inject enough energy to significantly accelerate the ejected material, thereby driving a fast shock through a large fraction of the ejecta (e.g., \citealt{Beloborodov2020}). This may even be the source of the fast ($\sim 0.3$ c) component in GW170817. Such energy injection can ensue before and/or after the  nucleosynthesis freeze-out and may have a significant effect on the composition of most of the ejecta
\subsection{Alpha particles}\label{sec:BNS_alpha}
 Estimates based on the observed kilonova emission in GW170817 indicate ejecta mass of  $M_{ej}\approx 10^{-2} \, M_\odot$.
The corresponding density at a radius $r= 10^{10} r_{10}$ cm is $\rho \approx 10 (M_{ej}/10^{-2} M_\odot)r_{10}^{-3}$ gr cm$^{-3}$.
The blackbody temperature in the far downstream, achieved when the downstream plasma reaches full thermodynamic equilibrium, is $kT_{d,BB} = 60 (\rho/10 \, \text{gr} \, \text{cm}^{-3})^{1/4}(\beta_u/0.3)^{1/2}$ keV. This is a strict lower limit on the temperature just behind the shock. In case of r-process rich upstream, \citet{LE2020} find that for a single-fluid RMS model, if no pairs are created, the immediate downstream falls out of equilibrium at $\beta_u>0.25 (\rho/10 \, \text{gr} \, \text{cm}^{-3})^{1/30}$, whereupon the temperature inside and just behind the shock is significantly higher than  $kT_{d,BB}$. We therefore conclude that for shocks with  $\beta_u>0.25$ that cross the ejecta within the first few seconds the  downstream temperature is expected to exceed $50$ keV, 
leading to excessive pair creation and, consequently,  a large abundance of positrons inside the shock \citep{katz2010,budnik2010,LE2020}. 
According to the results obtained in \ref{pairs_role}, such overabundance of positrons will lead to both, a reduction in the physical shock width 
and nearly complete screening of the electrostatic field inside the shock.
As a result, binary Coulomb collisions of pairs and ions, which under such conditions becomes the dominant coupling mechanism, is not strong enough to prevent  decoupling of the alpha particles from the heavy ions cluster within the shock transition layer.  
Then, collisions between neutron-rich isotopes and alpha particles just behind the shock may be strong enough to trigger fission and fusion. This can lead to a substantial 
composition change, provided anomalous friction does not prevent full decoupling inside the shock. 
Below, we provide an estimate of the scale separation 
(between radiation and kinetic scales) and discuss the implications of anomalous coupling.

\subsection{Free neutrons}\label{sec:BNS_neutrons} 
 As mentioned above, a large abundance of free neutrons is present in the ejecta up to about one second post-merger.
If the shock propagates in the ejecta during this stage, free neutrons advected with the upstream flow will cross the shock 
undisturbed and collide with the heavy nuclei downstream.  This can trigger fission, provided that
 the relative velocity of the free neutrons with respect to the 
downstream ions exceeds the fission barrier.   Fission barriers of heavy ions, at or near the valley of stability, range from a few MeV for atomic numbers $Z > 80$ roughly (depending on $A$),
to about 40 MeV for lighter elements \citep{moller2009,royer2021}.  For most ions with $Z >70$  it is roughly 20 MeV. According to the results in \ref{sec:free_neutrons}, the energy of free neutrons in the downstream exceeds this value at shock velocities $\beta_u\gtrsim 0.25$. Prior to freeze-out, all the nuclei are far from the valley of stability and, therefore, even lower energy neutrons are likely to cause transmutations of heavy nuclei. Finding the effect of the sudden heating of free neutrons by shocks, after nucleosynthesis starts to fall out of equilibrium and before freeze-out, on the final ejecta composition is out of the scope of this paper. Yet, we suspect it to be significant.




\subsection{Scale separation and coupling length}
\label{sec:scale_separation}

In section \ref{friction}, we demonstrated that sufficiently strong collective inter-ionic friction can tightly couple the ions,
preventing relative drifts, when binary collisions fail to do so. As explained there, in reality, such friction represents anomalous momentum transfer between ions via scattering off plasma turbulence.
Such momentum transfer occurs on microscopic scales, but the actual length scale is unclear at present. In \citet{Arno_plasma_filaments} 
it has been shown, by means of PIC simulations,  
that in a single-ion RRMS, where positrons are present, the anomalous coupling length scales as  $l_c \sim 10^5  {M}^{1/2} (\gamma_u/10)^{-1} l_p$,
where $M$ is the pair multiplicity, $\gamma_u$ the shock Lorentz factor and $l_p$ the proton skin depth defined below. For the typical
densities expected in many exploding stellar systems (e.g., SNe, LGRB) $l_c$ has been shown to be much smaller than the shock width (by several orders of magnitude).
This result is probably irrelevant to the Newtonian, multi-ion RMS studied here, as the plasma modes that dominate the turbulence are most likely 
different.  However, it suggests that the anomalous coupling length may encompass thousands or even tens of thousands of skin depths.
At the extreme densities of BNS merger ejecta, this may exceed the shock width, as we now show.

To get a rough measure of the anticipated scale separation in RMS, 
we compare the Thomson length, $l_T = (\sigma_T n_{e,u})^{-1}$, with the proton
skin depth at the same density, $l_p = c/\omega_p = (m_pc^2/4\pi e^2n_{e,u})^{1/2}$ \footnote{The skin depth of an individual ion beam $\alpha$ is related 
to the proton skin depth through: $l_\alpha = \sqrt{n_{e,u} A_\alpha/n_{\alpha,u}Z_\alpha^2}\, l_p \simeq l_p$.}.
The upstream electron number density can be estimated upon assuming that all elements have $Z/A = 0.5$:
\begin{equation}\label{eq:density_BNS}
    n_{e,u}\approx \frac{\rho}{2m_p} \approx 10^{25} (M_{ej}/10^{-2} M_\odot)r_{10}^{-3}  \quad  \text{cm}^{-3},
\end{equation}
from which we find 

\begin{equation}
    l_T/l_p \approx 10^{4} \left(\frac{n_{e,u}}{10^{25}\, \text{cm}^{-3}}\right)^{-1/2}.
\end{equation}
Note that this ratio scales as $r_{10}^{3/2}$.  Comparing the ratio $l_T/l_c$ in the last equation with the ion collision depth in Fig. \ref{fig:R_process_M10}
suggests that if indeed $l_c> 10^4 l_p$, anomalous coupling may not prevent inelastic ion-ion collisions, and the consequent change of the composition profile. 
A more firm conclusion requires a quantitative analysis of plasma instabilities in the multi-ion system, which is beyond the scope of this paper.  
We defer such analysis for a future study.


\section{Summary}

We have demonstrated how the passage of a fast radiation-mediated shock through a multi-ion plasma can lead to significant drift velocities between various plasma constituents. The reason is that ions having different charge-to-mass ratios experience different deceleration rates by the electrostatic
field generated inside the shock, owing to the charge separation imposed by the radiation force acting solely on the electrons. We have also shown that the presence of pairs and/or free neutrons enhance the expected drift velocities, while anomalous friction due to plasma instabilities is acting to reduce the drift velocities.

Drift velocities will, first, lead to the generation of micro-turbulence which can in turn lead to the acceleration of particles, thereby affecting the emission from such shocks upon breakout, as well as other properties of the shock. Most interestingly, for shocks that are fast enough the development of significant drift velocities can trigger nuclear reactions in the immediate shock downstream, thereby changing the composition behind the shock.

In case of BNS mergers, we find that nuclear reactions might be triggered by shocks that cross the ejecta during the first few seconds after the merger at velocities $\beta_u\gtrsim0.25$. If the shocks cross the ejecta before nucleosynthesis freeze-out in the unshocked ejecta (roughly 1 s after merger), the highly abundant free neutrons are heated to temperatures of tens of MeV (compared to sub-MeV temperatures before shock crossing). This process is not affected by anomalous friction. While we do not compute the effect of such heating on the final composition, we expect it to be significant.   
If fast shocks cross the eject after nucleosynthesis freeze-out, collisions of $\alpha$ particles with heavy ions in the immediate downstream are expected, possibly leading to nuclear transmutation. We have shown that the separation between radiation and kinetic scales in merger ejecta is small enough to possibly prevent anomalous friction from playing a significant role in these systems.
The resultant change in composition behind the shock that the jet drives as it crosses the ejecta, might considerably affect the properties of the early kilonova 
emission (minutes to hours), which is likely dominated by the cocoon. The late kilonova emission can be also affected if fast shocks arise between different components of the ejecta or if the central remnant injects a significant amount of energy into the outflow after its ejection. 

In SNe, if significant drift velocities are not prevented by anomalous friction, proton capture reactions are found to be important at shock velocity $\beta_u\gtrsim0.2$, even if the abundance of H is
smaller than solar by several orders of magnitudes.  Such high velocities are not expected in type II SNe, but are likely in type Ib/c SNe
and in shock breakout from extended stellar envelopes in llGRB \citep{nakar2015}. 
The absence of hydrogen lines in the spectra of these SNe sets a limit on the H abundance;  according to \citet{Hachinger2012} an H mass  fraction of even $1 \%$ out of the total
may be allowed, where the H mass fraction in the outer envelope can be considerably higher.  What would be the effect of the composition change on the emitted spectrum in these sources is unclear at present, although it would probably affect only spectra at very early times. 
We hope that future studies might be able to identify observational diagnostics that can test this model.

Our analysis ignores the effect of anomalous friction due to plasma instabilities.   An estimate of the ratio between kinetic and radiation scales is given in Sec. \ref{sec:scale_separation}.
Based on this estimate, we speculated that in BNS mergers, owing to the extreme density of the ejecta, anomalous coupling may not prevent 
a complete velocity separation between He and the heavy ions,  
hence, may not affect our results significantly.  In SNe, where the typical densities are 
smaller by about ten orders of magnitudes, the scale separation is much larger and the effect of beam instabilities on the downstream velocity 
distribution of the ions is unclear.  We point out that if the threshold drift velocity required for anomalous coupling is large enough, 
it may lead to formation of collisionless subshocks, with random ion velocities in excess of the proton capture barrier.   Otherwise, 
the onset of instabilities might suppress nuclear reactions in SNe.  We plan to conduct a detailed kinetic study of instabilities in multi-ion RMS,
along similar lines to the study described in \citet{Arno_plasma_filaments}, in the near future.

\section*{Data Availability}
The data underlying this article will be shared on reasonable request to the corresponding author.

\section*{Acknowledgements}
We thank Israel Mardor and Heinrich Wilsenach for enlightening discussions on nuclear physics.
This research was partially supported by an Israel Science Foundation grant 1995/21 and a consolidator ERC grant 818899 (JetNS) .

\bibliographystyle{mnras}
\bibliography{bib}

\appendix
\section{Derivation of shock equations}\label{Eq Dev}

We consider a planar shock propagating in the $-\hat{x}$ direction through a static, fully ionized, infinite cold medium. 
In the shock frame the flow is steady, with an upstream velocity $\beta_u\lesssim0.5$ in the $\hat{x}$ direction. 
We suppose that the plasma upstream of the shock is made of electrons and various ion species (henceforth labeled by a subscript $\alpha$)
of atomic number $Z_\alpha$ and mass number $A_\alpha$.  We further assume that the energy dissipated inside the shock is fully converted into radiation,
which is a good approximation in RMS.  For simplicity, we neglect pair production inside the shock (which may not be justified at extreme 
densities, see Sec. \ref{sec:BNS_mergers_applc} for details) and any other  conversion process (e.g., nuclear transmutations).
In the multi-fluid approach developed here, the electrons and the different ion species are treated as different fluid components, with densities $n_e$ and $n_\alpha$, respectively,
and velocities $\beta_e$ and $\beta_\alpha$.  Since we neglect conversion processes, all 
particle fluxes must be conserved across the shock, viz., $j_e \equiv n_e\beta_e =$ const, and likewise for $j_\alpha \equiv n_\alpha \beta_\alpha$.

Following \citet{BPb1981}, we invoke the diffusion approximation to compute the transfer of radiation through the shock.
Since, practically, only the electrons experience the radiation force (through inverse Compton scattering), the advection velocity
must equal $\beta_e$.  The energy flux of the radiation (in units of $c=\hbar=1$) can then be expressed as \citep{BPb1981,LE2020}:

\begin{equation}\label{eq:app_Tox}
    T^{0x}_\gamma=\beta_e (e_\gamma+p_\gamma) - \frac{1}{3n_e \sigma_T}\partial_x e_\gamma,
\end{equation}
where $e_\gamma$ and $p_\gamma$ are the radiation energy density and pressure, respectively. 
The first term on the R.H.S describes the advection of enthalpy, and the second term the diffusion flux (Fick's law), 
with the diffusion coefficient averaged over all angles (hence the $1/3$ factor). 
Treating the radiation as a relativistic gas, we can use the equation of state $e_\gamma=3p_\gamma$ to close the system of moment equations. 
With the normalization $\pi_\gamma = p_\gamma/ m_e j_e$ and $\tilde{T}^{0x}_\gamma= T^{0x}_\gamma /m_e j_e$, Eq. \eqref{eq:app_Tox}
reads:

\begin{equation}
\Tilde{T}^{0x}_\gamma = 4\pi_\gamma \beta_e -\partial_\tau \pi_\gamma.
\label{eq:radiation_E_flux}
\end{equation}

Under the assumption that the plasma is cold, the energy fluxes of the electron and ion fluids are given 
by $T^{0x}_e=\frac{1}{2}m_e n_e \beta_e ^3$ and $T^{0x}_\alpha=\frac{1}{2} A_\alpha m_p n_\alpha \beta_\alpha ^3$, respectively. 
 In terms of the Thomson optical depth, $d\tau=n_e \sigma_T dx$, energy conservation across the shock can be expressed as,
\begin{equation}\label{eq:app_E_cons}
 \partial_\tau (T^{0x}_\gamma+T^{0x}_e+\sum_\alpha T^{0x}_\alpha)=0.
\end{equation}

To the order we are working in, the  momentum fluxes of the radiation and the electron fluid are, respectively,  
$T^{xx}_\gamma= p_\gamma$ and $T^{xx}_e = m_e n_e \beta_e^2 = m_e j_e \beta_e$. The change in the momentum flux of the coupled 
electron-photon system must equal the  electrostatic force density acting on the electrons:
\begin{equation}\label{eq:app_momentum_e}
    \partial_x(T^{xx}_e + T^{xx}_\gamma) = -e n_e E.
\end{equation}
Dividing by $m_e j_e$ and using our normalization, we obtain Eq. \eqref{eq:1},
\begin{equation}\label{eq:app_mom_e_normal}
     \partial_\tau (\pi_\gamma+\beta_e) = - \Tilde{E}. 
\end{equation}

Likewise, the momentum flux of ion species $\alpha$ is $T^{xx}_\alpha = m_pA_\alpha n_\alpha \beta_\alpha^2 = m_pA_\alpha j_\alpha \beta_\alpha$.
The change in this momentum flux must equal the electrostatic force density acting on the ions:
\begin{equation}\label{eq:app_mom_ions}
    \partial_x T^{xx}_\alpha  = e Z_\alpha n_\alpha E.
\end{equation}
Dividing by $m_p A_\alpha  j_\alpha$ and normalizing we obtain Eq. \eqref{eq:2}.
To obtain Eq. \eqref{eq:4} we note that $\partial_x T_\alpha^{0x}= \beta_\alpha \partial_x T_\alpha^{xx}= e Z_\alpha j_\alpha E$.
Likewise, $\partial_x T_e^{0x}= \beta_e \partial_x T_e^{xx} = - e  j_e E -\beta_e\partial_x T_\gamma^{xx}$, where Eq. \eqref{eq:app_momentum_e} has been employed.

By summing over all ions and using  Eq. \eqref{eq:continuity}, we find
\begin{equation}\label{eq:app_E_P}
  \partial_x( T^{0x}_e+\sum_\alpha T^{0x}_\alpha )=   -\beta_e\partial_x T_\gamma^{xx} = - \beta_e\partial_x p_\gamma.
\end{equation}
Combining the last equation with Eqs. \eqref{eq:radiation_E_flux}-\eqref{eq:app_E_cons},
we  finally arrive at Eq. \eqref{eq:4}:
\begin{equation}\label{eq:app_E_rad}
    \partial_\tau \Tilde{T}^{0x}_\gamma =\partial_\tau ( 4\pi_\gamma \beta_e -\partial_\tau \pi_\gamma) =\beta_e \partial_\tau \pi_\gamma.
\end{equation}

\subsection{Inclusion of interspecies friction} 
\label{app_firc_derivation}
We suppose that the change in momentum of species $a$ ($a=e$ for electrons and $a=\alpha$ for ions) due to momentum exchange with species $a'$ is
proportional to the relative velocity between the two beams and the density of the $a'$ beam: $\mathcal{F}_{a a'}=g_{a a'} n_{a'} (\beta_a-\beta_{a'})$, 
where  $g_{a a'}$ is a characteristic friction coefficient that, in general, can depend on the velocities and temperatures of the colliding fluids.
With friction forces included, the momentum equations of the electrons and ions read:
\begin{align}
      & \partial_x(T^{xx}_e + T^{xx}_\gamma) = -e n_e E + \sum_\alpha n_e\mathcal{F}_{e \alpha}, \label{eq:app_momentum_e_fric} \\
   & \partial_x T^{xx}_\alpha  = e Z_\alpha n_\alpha E + \sum_{\alpha'} n_\alpha\mathcal{F}_{\alpha \alpha'} + n_\alpha \mathcal{F}_{\alpha e}.
\end{align}
The anti-symmetry of the interspecies forces, $n_a \mathcal{F}_{a a'} + n_{a'}\mathcal{F}_{a' a} = 0$, guarantees momentum conservation
of the entire system (plasma and radiation).  However, since friction forces are dissipative (non-conservative), additional physics is
needed to specify how the dissipation energy is distributed.   Here, we adopt a simplified treatment that invokes ad hoc energy transfer between ions 
and electrons over length scales much shorter than the radiation length.  Specifically, we assume that the electron and  ions  fluids remain highly supersonic 
at all times.   Under this assumption the energy fluxes of the electrons and ions can still be expressed as 
$T^{0x}_e=\frac{1}{2}m_e n_e \beta_e ^3$ and $T^{0x}_\alpha=\frac{1}{2} A_\alpha m_p n_\alpha \beta_\alpha ^3$, respectively, whereby
$\partial_x T_a^{0x}= \beta_a \partial_x T_a^{xx}$ for $a \in [e,\alpha]$.   Energy conservation, Eq. \eqref{eq:app_E_cons}, then yields
\begin{equation}\label{eq:app_energ_fric}
     \partial_x \Tilde{T}^{0x}_\gamma =  - \sum_a \beta_a \partial_x \Tilde{T}^{xx}_a
     = \beta_e  \partial_x p_\gamma - \sum_{a,a'} n_a\beta_a \, \mathcal{F}_{aa'}, \quad a\in [e,\alpha] .
\end{equation}


\subsection{Inclusion of positrons}
\label{app_sec_pairs}
We now re-derive the equations assuming that a fixed number of positrons is contained the upstream flow. We denote the density of positrons 
by $n_+$ and the total density of electron by $n_e$ as before.  The corresponding conserved fluxes are $j_+ = n_+ \beta_+$ and $j_e = n_e\beta_e$.  
The pair multiplicity is defined as $M = n_{+,u}/(n_{e,u}-n_{+,u})$, noting that 
$n_e-n_+ = \sum_\alpha n_\alpha Z_\alpha$ is the  total number of protons in the system. The neutrality condition now reads:
\begin{equation}\label{eq:pairs_flux_cons}
    \sum_{\alpha}j_{\alpha}Z_{\alpha}+j_{+}=j_{e}.
\end{equation}
We normalize all quantities as before in terms of the total electron flux $n_e$ and flux $j_e$.  The optical is then $d\tau = \sigma_T n_e dx$, and the fiducial 
electric field $E_0$, the parameter $\chi$ and the radiation pressure are expressed as before in therms of $j_e$. 
With positrons included, Gauss' equation, Eq. \eqref{eq:Gauss}, becomes 
 \begin{equation}\label{eq:pairs_E_pos}
 \partial_\tau \tilde{E}=\chi\left[-1 +\frac{M \beta_{-}}{(1+M)\beta_{+}}+\sum\frac{\beta_{-}}{\beta_{\alpha}}\frac{j_{\alpha}}{j_{e}}Z_{\alpha}\right].  
\end{equation}

Since the radiation is coupled to positrons in addition to electrons, the advection of radiation in the shock depends on both $\beta_e$ and $\beta_+$, and
Eq. \eqref{eq:app_Tox} is no longer valid.  We settle for a simple prescription, defining 

\begin{equation}
\Tilde{T}^{0x}_\gamma = 4\pi_\gamma\left( \frac{n_e \beta_e}{n_e+n_+} + \frac{n_+ \beta_+}{n_e+n_+}\right) - \frac{n_e}{n_e+n_+}\partial_\tau \pi_\gamma.
\label{eq:radiation_E_flux_pos}
\end{equation}


The momentum flux of the radiation remains $T_\gamma^{xx} = p_\gamma$.  Now, the momentum transferred to the electron 
fluid by the radiation at any given position is proportional to the relative number of scatterings, $n_e/(n_e+n_+)$, and 
likewise for the positron fluid. Consequently, the momentum equations for the electron and positron fluids are, respectively, 
\begin{equation}
\begin{split}\label{eq:app_momentum_ep}
 \partial_x \left(T_e^{xx} + \frac{n_e}{n_e+n_+}p_\gamma \right) & = - en_e E  + n_e \mathcal{F}_{e+} + \sum_\alpha n_e\mathcal{F}_{e \alpha}  ,\\
  \partial_x \left(T_+^{xx} + \frac{n_+}{n_e+n_+}p_\gamma \right) & =  en_+ E  + n_+ \mathcal{F}_{+e} + \sum_\alpha n_+\mathcal{F}_{+ \alpha}.
\end{split} 
\end{equation}
The ion momentum equation reads:
\begin{equation}\label{eq:app_momentum_ion}
    \partial_x T^{xx}_\alpha  = e Z_\alpha n_\alpha E + \sum_{\alpha'} n_\alpha\mathcal{F}_{\alpha \alpha'} + n_\alpha \mathcal{F}_{\alpha e} + n_\alpha \mathcal{F}_{\alpha +} .
\end{equation}
It is readily seen that the sum gives
\begin{equation}
\begin{split}
 \partial_x \left(T_e^{xx} + T_+^{xx} +\sum_\alpha T_\alpha^{xx}  + p_\gamma \right) = 0,
\end{split} 
\end{equation}
as required from momentum conservation.  

From energy conservation, $\partial_\tau (T^{0x}_\gamma+T^{0x}_e + T_+^{0x} +\sum_\alpha T^{0x}_\alpha)=0$, and the neutrality condition, Eq. \eqref{eq:pairs_flux_cons}, 
we finally arrive at
\begin{equation}
\begin{split}
    \partial_x T^{0x}_\gamma & = \beta_e \partial_x \left(\frac{n_e}{n_e+n_+} p_\gamma \right) 
    +  \beta_+ \partial_x \left(\frac{n_+}{n_e+n_+} p_\gamma \right) 
    - \sum_{a,a'} n_a\beta_a \, \mathcal{F}_{aa'},
    \end{split}
\end{equation}
here $a\in [e,+,\alpha]$.
Note that if the pairs are strongly coupled, $\beta_e = \beta_+$, the sum of the first two terms on the R.H.S. reduces to $\beta_e \partial_x p_\gamma$,
as expected.





\section{Derivation of Coulomb friction coefficients}\label{coul_fric_eq}
We consider a gas of particles of type $a$ with bulk velocity $\beta_a c$ and number density $n_a$ moving into a gas of $a'$ particles. $a$ and $a'$ can either be an ion isotope ($_Z^A \alpha$) or electrons (with subscript $e$, taking the values $Z_e=-1$ and $A_e=\mu\equiv\frac{m_e}{m_p}$). 
The general force per unit volume acting on the gas of type $a$ moving through a gas of $a'$ particles is:
\begin{equation}\label{col_el_force}
\mathcal{F}_{aa'}=-m_a n_a \nu_{aa'}\Delta v_{drift,aa'}.
\end{equation}
Where $\nu_{aa'}$ is the coulomb interaction rate given by:
\begin{equation}
\nu_{aa'}=\sqrt{2\pi}\frac{4}{3}n_{a'}\frac{Z_a^2Z_{a'}^2 e^4 (A_a+A_{a'})}{A_a^2 A_{a'} m_p^2 \Delta \beta_{aa'}^3 c^3} \ln{\Lambda_{aa'}}.
\end{equation}

The logarithmic factor $\Lambda_{aa'}$ comes out of integrating between minimal and maximal impact parameters. We used $r_{min}=\max\{\frac{4\pi Z_a Z_{a'}e^2}{A_a A_{a'} \Delta\beta_{aa'}^2 m_p c^2}(A_a+A_{a'}),\frac{\hbar c}{A_a A_{a'} \Delta\beta_{aa'} m_p c}(A_a+A_{a'})\}$ and $r_{max}=\lambda_{debye}=\sqrt{\frac{\theta_e m_e c^2}{4\pi n_e e^2}}$. All together one finds:
\begin{equation}\label{eq:app_col-log}
\begin{split}
\Lambda_{aa'}&=\frac{r_{max}}{r_{min}}=\\
&=\frac{1}{\mu(4\pi)^{3/2}}\cdot\frac{(m_e c^2)^{3/2}}{(e^2)^{3/2}}\cdot\frac{A_a A_{a'}}{Z_a Z_{a'}(A_a+A_{a'})}\cdot\frac{\theta_e^{1/2}}{n_e^{1/2}}\Delta\beta_{aa'}^2=
\\&=2.6\cdot10^{20}\cdot\frac{A_a A_{a'}}{Z_a Z_{a'}(A_a+A_{a'})}\cdot\frac{\theta_e^{1/2}}{n_e^{1/2}}\Delta\beta_{aa'}^2.
\end{split}
\end{equation}
\\
In case $r_{min}>r_{max}$ the interaction is totally screened and we set $\Lambda_{a a'}\equiv1$ so that $\log\Lambda_{a a'}=0$. Regarding $\Delta\beta_{aa'}^2$ one has to take into account both the relative drift and thermal velocity. We obtain:
\begin{equation}
\begin{split}
\Delta\beta_{aa'}^2=\Delta\beta_{drift,aa'}^2&+\Delta\beta_{thermal,aa'}^2=
\\&=(\beta_a-\beta_{a'})^2+2\mu(\frac{\theta_a}{A_a}+\frac{\theta_{a'}}{A_{a'}}).
\end{split}
\end{equation}
Writing down the force equation in the shock parameters, $\mathcal{F}_{aa'}=A_a m_p n_a \beta_a n_e \sigma_T \frac{d\beta_a}{d\tau} c^2$, combining \ref{col_el_force},  one finds:
\begin{equation}\label{ion_ion_col_force}
\frac{d\beta_a}{d\tau}=-\frac{\beta_e}{\beta_a A_a} \frac{\Tilde{j}_{a'}}{\beta_{a'}}\frac{\mu^2}{\sqrt{2\pi}}\frac{A_a+A_{a'}}{A_a A_{a'}}Z_a^2Z_{a'}^2 \frac{\ln{\Lambda_{aa'}}}{\Delta\beta_{aa'}^3} \Delta\beta_{drift,aa'}
\end{equation}
where we used $\sigma_T=\frac{8\pi}{3}\cdot\frac{e^4}{m_e^2 c^4}$, $\mu\equiv\frac{m_e}{m_p}$, $\Tilde{j}_{a'}\equiv\frac{n_{a'}\beta_{a'}}{n_e\beta_e}$ and $d\tau\equiv n_e \sigma_T dx$. Using previous notations we find:
\begin{equation}
    \Tilde{g}_{aa'}=\frac{\mu^2}{\sqrt{2\pi}}\frac{A_a+A_{a'}}{A_a A_{a'}}Z_a^2Z_{a'}^2 \frac{\ln{\Lambda_{aa'}}}{\Delta\beta_{aa'}^3}.
\end{equation}

To estimate the conditions under which relative drift would rise between the particles, we need to find when the thermal-dominated coupling breaks down. In case the interaction is temperature-dominated, assuming all particles own the same temperature $\theta_a=\theta_{a'}\equiv\theta$, one obtains $\Delta\beta_{aa'}^2\approx2\mu\theta \frac{A_a+A_{a'}}{A_a A_{a'}}$.
This gives:
\begin{equation}
 \Tilde{g}_{aa'}=\sqrt{\frac{\mu}{16\pi}\cdot\frac{A_a A_{a'}}{A_a+ A_{a'}}}Z_a^2Z_{a'}^2 \frac{\ln{\Lambda_{aa'}}}{\theta^{3/2}}
\end{equation}
with $\Lambda_{aa'}=5.2\cdot10^{20} \frac{\mu}{Z_a Z_{a'}}\frac{\theta^{3/2}}{n_e^{1/2}}$.\\
Requiring complete ion-ion coupling is done by taking $\Delta\beta_{drift,aa'}=d\beta_a$ in \ref{ion_ion_col_force}. This enables us to find a relation between the length scale for inter-ionic coulomb friction coupling and the shock width $l_{sh}\equiv(\beta_u n_e \sigma_T)^{-1}$:
\begin{equation}\label{eq:coul_friction_length}
\begin{split}
l_{aa'}&= \sqrt{\frac{16\pi}{\mu}\frac{A_a +A_{a'}}{A_a A_{a'}}}\beta_u^2 A_a \frac{n_e}{n_{a'}} \frac{\theta^{3/2}}{Z_a^2 Z_{a'}^2 \ln{\Lambda_{aa'}}} l_{sh} 
\\ &\approx 3\cdot10^2 \beta_u^2 \sqrt{\frac{A_a +A_{a'}}{A_a A_{a'}}}\frac{A_a}{Z_a} \frac{n_e}{Z_{a'} n_{a'}} \frac{\theta^{3/2}}{Z_a Z_{a'} \ln{\Lambda_{aa'}}} l_{sh} ,
\end{split}
\end{equation}
where we assume $\Delta\beta_{aa'}^2 \propto \theta$, implying that this equation is valid as long as the thermal velocity is larger than the drift velocity.
When $l_{aa'}$ exceeds $l_{sh}$, the different constituents can no longer be coupled within the shock, and relative drift arises. 
Taking $a=\alpha$ and $a'=e$, \ref{eq:coul_friction_length} reduces to 
\begin{equation}\label{eq:e_coul_fric_length}
l_{\alpha e}= 1.3\cdot10^4 \frac{\theta_e^{3/2}}{\ln{\Lambda_{\alpha e}}} \frac{\beta_u^2 A_\alpha}{Z_\alpha ^2} l_{sh}. 
\end{equation}

\bsp	
\label{lastpage}
\end{document}